\documentclass[preprint2]{aastex}
\usepackage{epsfig}
\usepackage{natbib}

\begin{document}
\newcommand{\kms}{km~s$^{-1}$}
\newcommand{\Msun}{M_{\odot}}
\newcommand{\Lsun}{L_{\odot}}
\newcommand{\ML}{M_{\odot}/L_{\odot}}
\newcommand{\etal}{{et al.}\ }
\newcommand{\hhh}{h_{100}}
\newcommand{\hsq}{h_{100}^{-2}}
\newcommand{\tn}{\tablenotemark}
\newcommand{\mdot}{\dot{M}}
\newcommand{\p}{^\prime}
\newcommand{\kmsMpc}{km~s$^{-1}$~Mpc$^{-1}$}

\title{Galaxy Groups}

\author{R. Brent Tully,}
\affil{Institute for Astronomy, University of Hawaii, 2680 Woodlawn Drive,
 Honolulu, HI 96822, USA}

\begin{abstract}
Galaxy groups can be characterized by the radius of decoupling from cosmic expansion, the radius of the caustic of second turnaround, and the velocity dispersion of galaxies within this latter radius.  These parameters can be a challenge to measure, especially for small groups with few members.  In this study, results are gathered pertaining to particularly well studied groups over four decades in group mass.  Scaling relations anticipated from theory are demonstrated and coefficients of the relationships are specified.  There is an update of the relationship between light and mass for groups, confirming that groups with mass of a few times $10^{12} \Msun$ are the most lit up while groups with more and less mass are darker.  It is demonstrated that there is an interesting one-to-one correlation between the number of dwarf satellites in a group and the group mass.  There is the suggestion that small variations in the slope of the luminosity function in groups are caused by the degree of depletion of intermediate luminosity systems rather than variations in the number per unit mass of dwarfs.  
Finally, returning to the characteristic radii of groups, the ratio of first to second turnaround depends on the dark matter and dark energy content of the universe and a crude estimate can be made from the current observations of $\Omega_{matter} \sim 0.15$ in a flat topology, with a 68\% probability of being less than 0.44.

\end{abstract}

\section{Introduction}

In the imaginary world of simulations, researchers have a well developed picture of the collapse of matter into halos.  Over time, small halos are absorbed into larger units.  Collapsed regions filled with substructure can be defined by hundreds or thousands of particles.  Halos can be identified with precision within those simulations \citep{2009MNRAS.398.1150B}.

In the real world, most galaxies are observed to lie in groups or clusters.  However, membership may be so limited that the structure is ill defined.  There is no concensus among observers about what is meant by the terms `group' and `cluster'.  For example, it is commonly accepted that we live in something called the Local Group.  A quantitative boundary that would include the traditional members is the zero--velocity surface  \citep{2009MNRAS.393.1265K}.  Galaxies inside this surface are infalling or on more complex bound orbits.  Galaxies outside this surface are participating in the Hubble expansion.  But now consider the Virgo Cluster.  This entity would traditionally be taken to include the region within 2 Mpc where several thousand galaxies are following randomized orbits \citep{1987AJ.....94..251B}. The zero--velocity surface around the Virgo Cluster lies at a radius of 7.2 Mpc \citep{2014ApJ...782....4K}, almost half of our distance of 16.4 Mpc from the cluster.  The implicit traditional definitions of the Local Group and the Virgo Cluster are inconsistent.

It does not seem useful to draw a distinction between the terms `cluster' and `group'.  Clusters contain a lot of galaxies and groups contain only a few, but there is no clear demarcation between the two.  In this article, the two terms will be used interchangeably.  

So how should an observer define a group?   Structure forms as sufficient matter accumulates through gravitational attraction to decouple from the expansion of the universe.  The matter falls together on nearly radial orbits and, given enough time, evolves toward dynamic equilibrium.  Two natural dimensions to describe a collapsed structure are the gravitational radius, $r_g$ and the radius at 200 times the `critical' density for a matter-dominated closed universe, $r_{200}$.  Here, lower case $r$ implies a 3-dimensional radius and an upper case $R$ will be used to denote a projected radius.  Accordingly, if all galaxies in a group sample are given equal weight (ie, galaxies of whatever luminosity are considered to be test particles within an environment dominated by distributed dark matter) then the gravitational radius is defined as 
\begin{equation}
R_g = {{\rm N}^2 \over \sum_{i<j} 1 / R_{ij}}
\label{rg}
\end{equation}
where $R_{ij}$ is the projected distance between pairs, and a virial mass estimate is given for the group by
\begin{equation}
M_{\rm v} = \alpha \sigma_p^2 r_g / G = (\pi/2) \alpha \sigma_p^2 R_g / G
\label{eq:Mv}
\end{equation}
where the line-of-sight velocity dispersion for the sample of N galaxies is
\begin{equation}
\sigma_p = \sqrt{\sum_i (v_i - <v>)^2 / {\rm N}} 
\end{equation}
given a group mean velocity $<v>$.  The parameter $\alpha$ equals 3 if orbits are isotropically distributed with an isothermal distribution, 2.6 with isotropic velocities with an NFW \citep{1997ApJ...490..493N} mass profile \citep{2007A&A...475..169M}, or 2.4 with an anisotropy model \citep{2005MNRAS.363..705M} that is a good fit to $\Lambda$CDM halos \citep{2010A&A...520A..30M}.

The alternative dimension $r_{200}$ is defined to coincide with a density $200 \rho_{crit}$ where $\rho_{crit} = 3 {\rm H}_0^2 / 8 \pi G$ at redshift zero.  This parameter can be determined in simulations with large numbers of test particles but it is not such a useful construct in the context of observations of small groups.  However,  \citet{1999ApJ...518...69M} calculate a ratio between $r_{200}$ and $r_g$ assuming $M(r) \propto r$ that can be reduced to the relationship (at $z=0$)
\begin{equation}
r_{200} = {\sqrt{\alpha} \sigma_p \over 10 {\rm H}_0} ~.
\end{equation} 

There may be other dimensions that are observationally useful.
Consider the collapse of a spherically symmetric overdense region in the expanding universe \citep{1972ApJ...176....1G,1984ApJ...281....1F,1985ApJS...58...39B}.  The time of collapse, $t_c$, depends on density, $\rho$: $t_c \propto \rho^{-1/2}$.  In the approximation of spherical collapse, all across the universe at
$t_c = today$ overdense regions with a common density will be separating from the cosmic expansion, creating zero--velocity surfaces.  The radius that encloses one of these regions will be called the first turnaround radius, $r_{1t}$.

At earlier times, regions of successively higher density collapsed.  There was a time and a corresponding density that lead to collapse and re-expansion to a pause before recollapse.  This pause and turnaround creates a caustic  \citep{1989RvMP...61..185S} at what will be called the second turnaround radius, $r_{2t}$.

At yet earlier times, regions of yet higher density collapsed and resulted in caustics of higher order turnarounds.  In the real world, departures from spherical symmetry would result in violent relaxation  \citep{1967MNRAS.136..101L}, departures from radial orbits, and blurring of the caustics.  The collapsed region will begin to evolve toward energy equipartition.  Such regions will be referred to as `quasi-virialized'. 

These theoretical musings provide a basis for a definition of groups.  It will be argued that radii of second turnaround and, more problematically of first turnaround, are observable.

The caustic of second turnaround for a group can potentially be defined by two features: a density discontinuity and a velocity dispersion discontinuity.  These features were displayed by  \citet{1985ApJS...58...39B} in the simple case of spherical infall of an initial top hat density excess.  In this simple case, density falls off as a power law with radius within $r_{2t}$ and velocity dispersions are large.  Beyond $r_{2t}$, the density is expected to take a downward step and velocities locally are expected to manifest coherent infall with modest dispersion.  The same salient features are recovered with N-body simulations based on the same spherical infall model \citep{2009MNRAS.400.2174V}.
These features might be hard to identify in observational information.  The density step might be obscured by noise from line-of-sight contamination.  The step from large to small velocity dispersions can easily be confused by the limitation of access to only line-of-sight velocities and lack of precise distances that would distinguish backside from frontside infall.  The situation with many groups will be messy but there are relatively clean cases where the desired information can be recovered.  For example one can identify the envelope in redshift--radius plots for rich clusters that contains gravitationally bound orbits \citep{1997ApJ...481..633D, 2003AJ....126.2152R}.  The challenge given attention here is to find analogous features in groups with much smaller populations.  The challenge can be met if the entities are sufficiently isolated or so nearby that their three dimensional structure can be discerned. 

The musings suggest that there should be interesting scaling relations between groups.  In the spherical approximation, the density at $r_{2t}$ is the same for all groups at a given epoch
\begin{equation}
{M(r_{2t}) \over r_{2t}^3} = {\rm constant}.
\end{equation} 
With the approximation $M(r) \propto r$ 
\begin{equation}
M(r_{2t}) = M_{\rm v} (r_{2t} / r_g)
\end{equation} 
so
\begin{equation}
{M_{\rm v} \over r_{2t}^3} \cdot {r_{2t} \over r_g} = {\alpha \sigma_p^2 \over G r_{2t}^2} = {\rm constant}.
\end{equation} 
Hence
\begin{equation}
r_{2t} \propto \sigma_p \propto M_{\rm v}^{1/3}.
\end{equation}
In subsequent sections, data will be presented that confirm these relationships and provide correlation zero points.

There is the prospect that equivalent scaling relationships can be found based on the first turnaround radius, $r_{1t}$.  The interest here is that the ratio $r_{1t}/r_{2t}$ is dependent on the dark energy content of the universe.  Collapse conditions at a given time depend almost exclusively on the matter density but dark energy affects the clock, hence the timing between collapse events.  See the discussion in Section 6.  Gary Mamon claims (unpublished) that in a flat universe with 70\% of the cosmological density in vacuum energy then $r_{1t}/r_{2t} = 3.3$ whereas in a flat universe with the critical density in matter then $r_{1t}/r_{2t} = 3.7$. 

Finally it is to be entertained if there are observational manifestations of the `zero gravity' surface at a radius $r_{zg}$ around a halo.  Only material within this surface will ever participate in collapse onto the halo.  It has been suggested \citep{2009A&A...507.1271C} that the region of the zero gravity surface is expected to be evacuated and can be identified by this property.  The importance of this feature depends on the properties of dark energy.

Unless otherwise stated, distances quoted in this paper are from the {\it Cosmicflows-2} compendium \citep{2013AJ....146...86T}.  These distances are compatible with the Hubble Constant value 75~\kmsMpc\ \citep{2012ApJ...758L..12S}.

\section{From Big to Small}

The questions to be addressed are whether there are observational features that distinguish the infall and quasi-virialized regions of halos and, if so, how these features scale with halo mass.  Most of the discussion in this paper will focus on modest to very small groups of galaxies, the markers for the overwhelming majority of identifiable halos.   The sites of intermediate mass halos, those in the range $5 \cdot 10^{12} - 10^{14} ~\Msun$, are investigated with a wide field imaging survey at the Canada-France-Hawaii Telescope (CFHT), initially using the 0.3 sq. deg. 12K CCD camera and then the 1 sq. deg. Megacam detector.    Follow up spectroscopy was undertaken with Subaru and Keck telescopes.   The study is extended to very small halos in the range $10^{11} - 5 \cdot 10^{12}~\Msun$ by giving consideration to the region within 4~Mpc where, outside the zone of obscuration, almost every galaxy brighter than $M_B = -11$ has probably been identified \citep{2004AJ....127.2031K}.  Almost all of these nearby galaxies have now been sufficiently observed with Hubble Space Telescope (HST) that an accurate distance is available from a measurement of the luminosity of the tip of the red giant branch, the TRGB method \citep{2009AJ....138..332J}. There is now detailed information on the grouping properties of galaxies down to the scales of associations of dwarfs \citep{2006AJ....132..729T}.  However, before giving attention to those intermediate and low mass structures the discussion will begin with a few words about a couple of large and familiar systems.

\subsection{Two Massive Clusters}

\noindent {\it The Coma Cluster}, at a distance of 100 Mpc,
is the nearest high latitude example of a rich and evolved cluster of galaxies.   A binary of two dominant ellipticals lies at the center and most galaxies in the core are early morphological types.  X-ray emission arising from hot intracluster gas is relatively uniformly distributed, indicative of a relaxed system, although there is the clear marker of the infall of a sub-unit with hot gas around NGC 4839, $45^{\prime} = 1.3$~Mpc from the center \citep{2003A&A...400..811N}.  Among the many contributions to velocity field studies are those by \citet{1999ApJ...517L..23G}, \citet{2003AJ....126.2152R}, and \citet{2002ApJ...567..178E}.  The latter reference identifies a small difference in velocity dispersion between giant and dwarf galaxies, separated at $M_b = -17$: $\sigma_{giant} = 979 \pm 30$ \kms\ and $\sigma_{dwarf} = 1096 \pm 45$~\kms.   Extending to a more extreme range of low surface brightness dwarfs with $-16.5 < M_R < -12.5$, Chiboucas et al. (2010) find a dispersion $\sigma_{LSB} = 1269 \pm 126$ \kms\ (but confined to the center of the cluster!).  Each of these differences is at the marginal level of $2.2 \sigma$.

\begin{figure}[htbp]
\begin{center}
\includegraphics[scale=0.4]{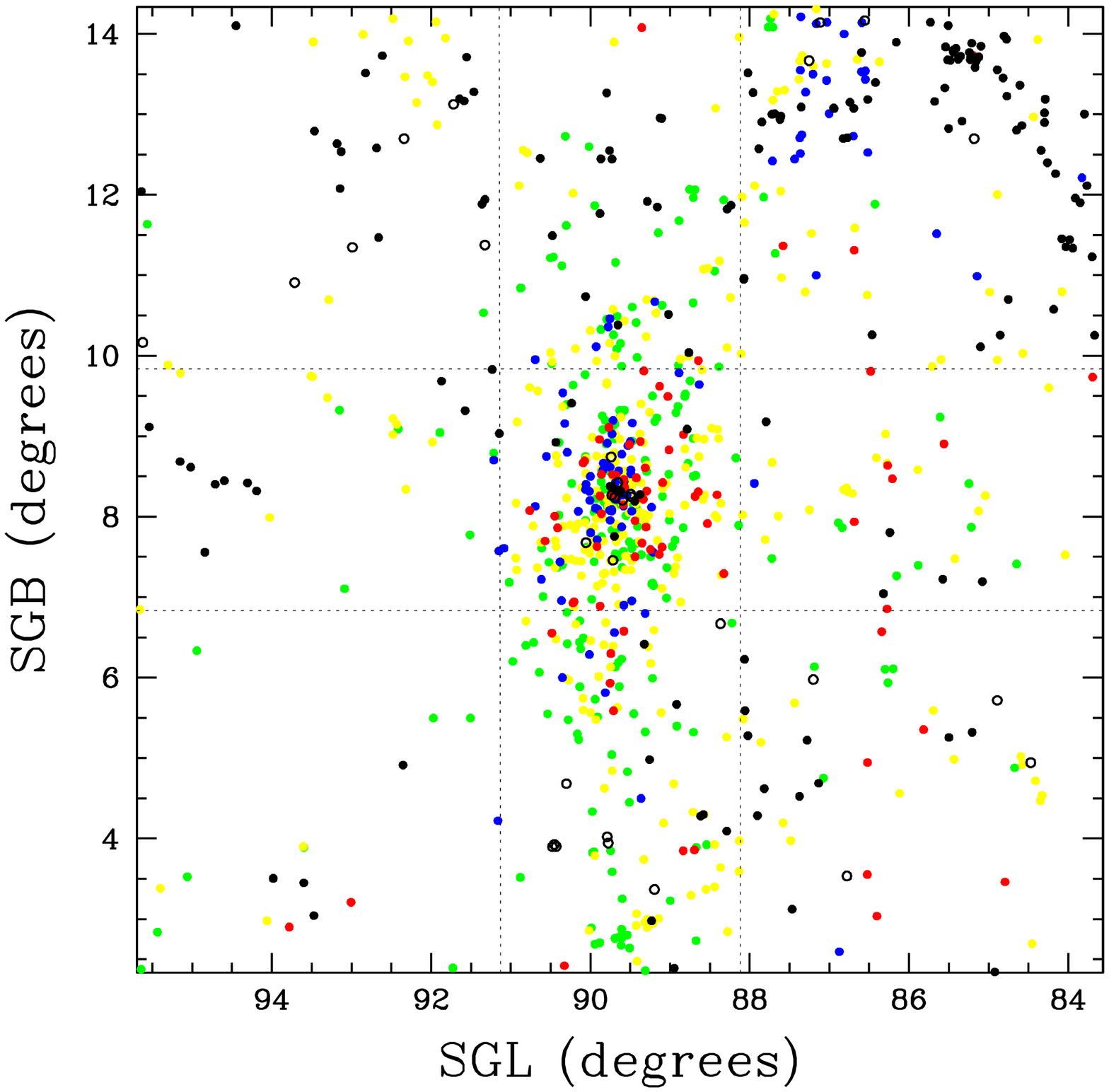}
\includegraphics[scale=0.4]{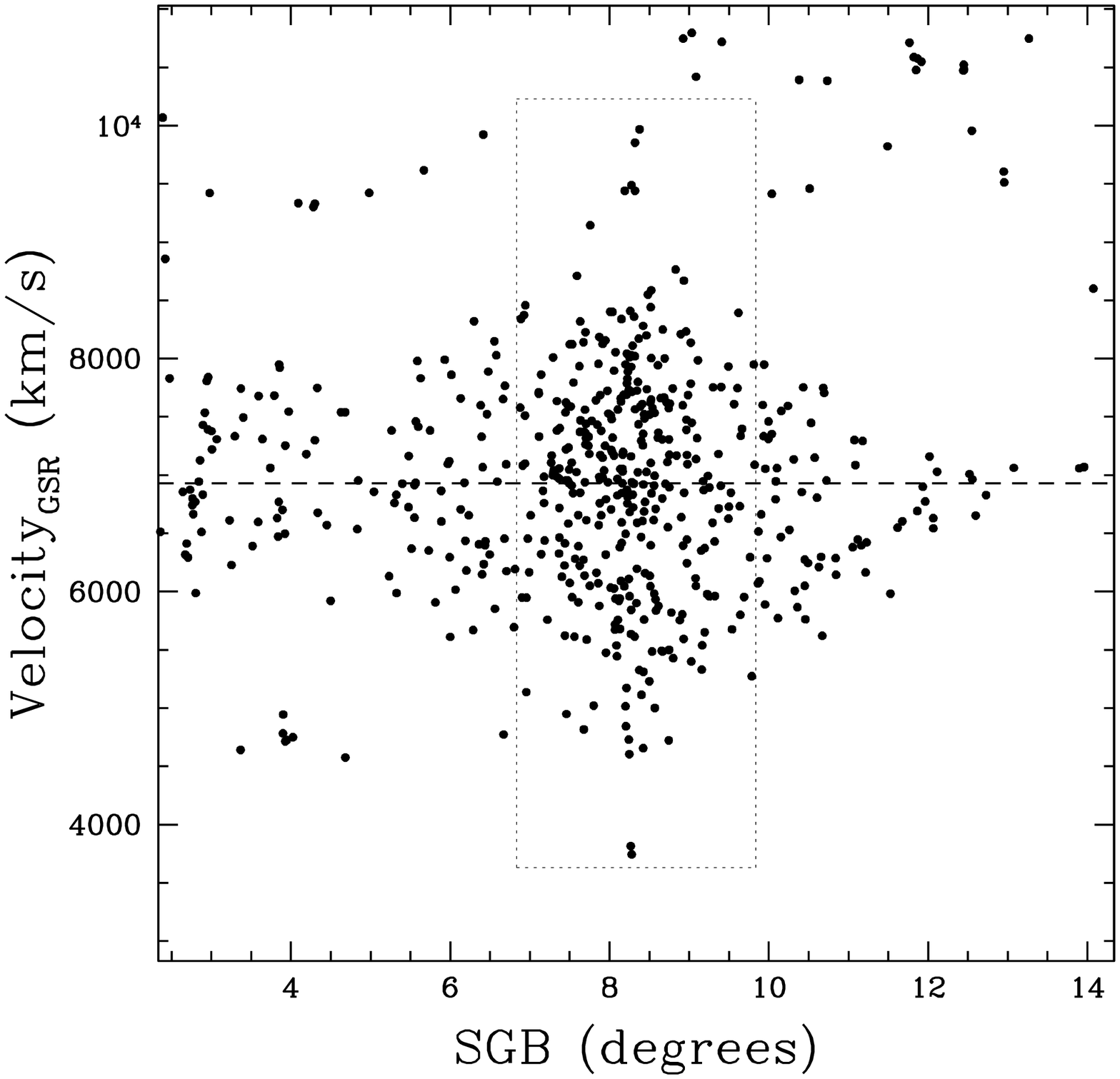}
\includegraphics[scale=0.4]{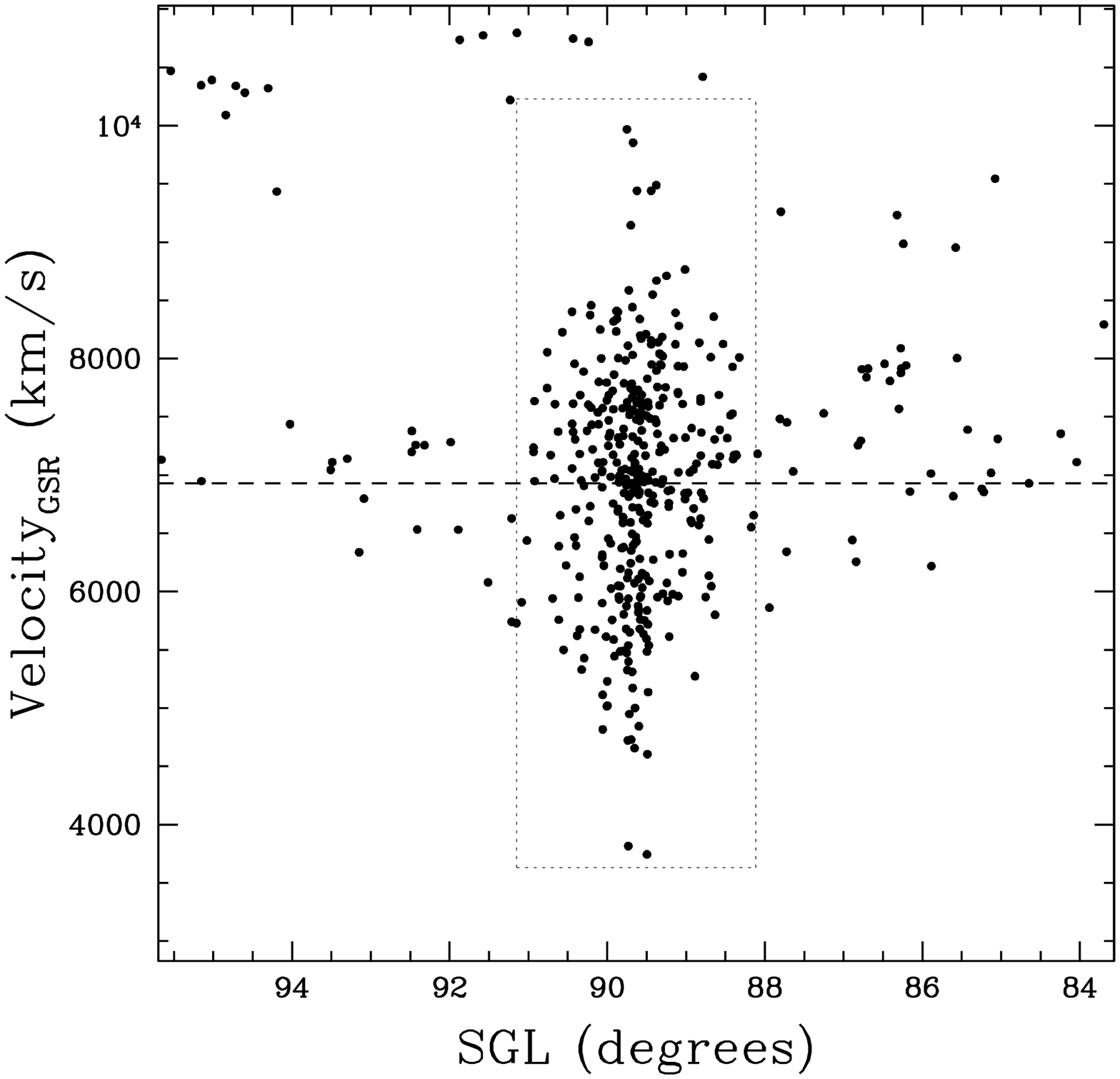}
\includegraphics[scale=0.4]{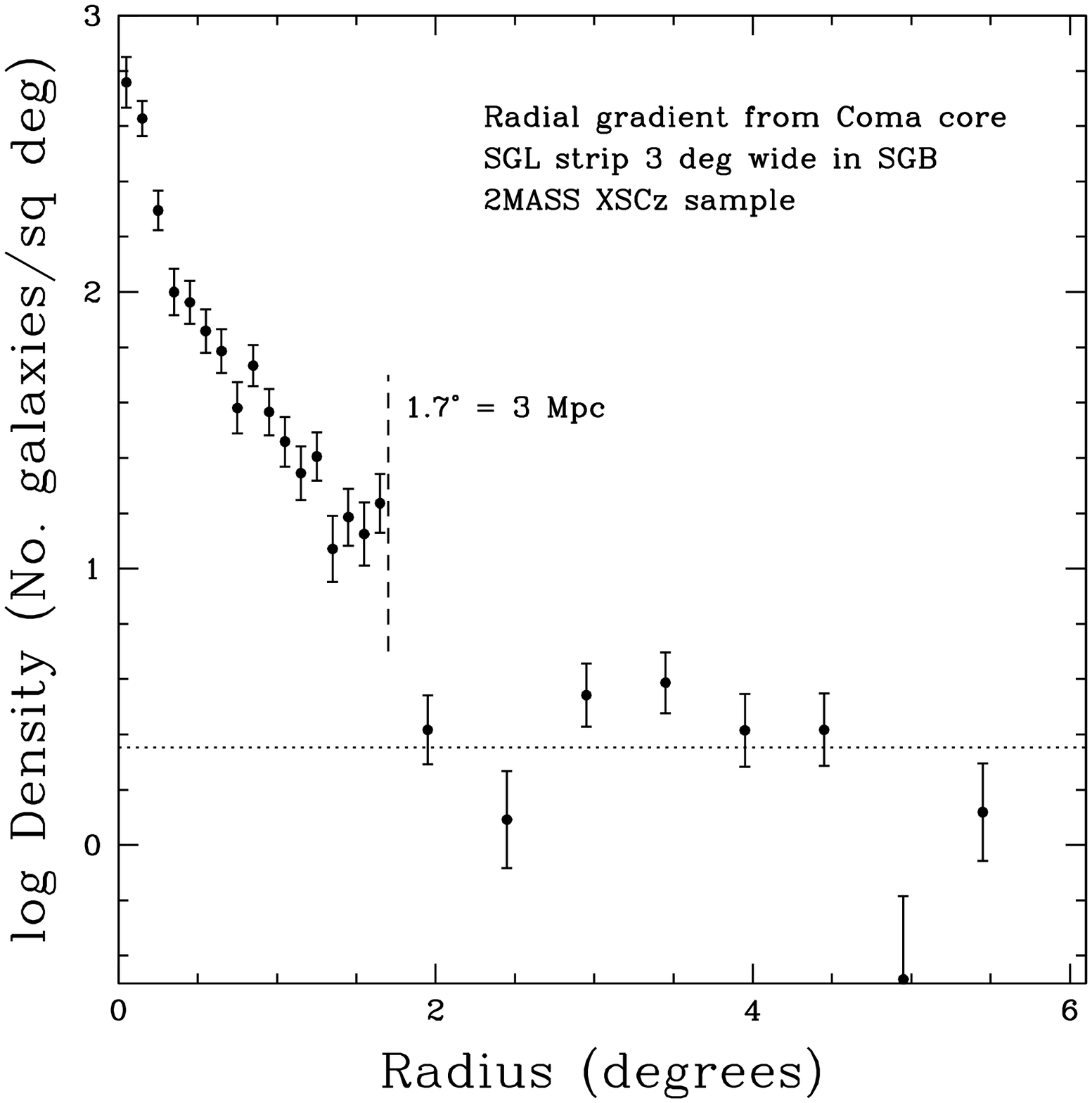}

\caption{{\it Top left:} Distribution of galaxies in the vicinity of the Coma Cluster in supergalactic coordinates.   Colors are coded by increasing redshift in 1000 \kms\ intervals from blue through red to black.
{\it Top right:} Velocities of galaxies within the central $3^{\circ}$ band in SGL of the top left figure as a function of SGB.   
{\it Bottom left:} Velocities of galaxies within the central $3^{\circ}$ band in SGB as a function of SGL.  The dotted box in this and the top right panel enclose galaxies within the collapsed Coma Cluster.   
{\it Bottom right:} Radial number density gradient in SGL from the Coma Cluster core in the central $3^{\circ}$ band in SGB.  
}
\label{coma}
\end{center}
\end{figure}

Our interest is in exploring the boundaries of the cluster so we give attention to a velocity survey that is complete over a wide angle.  We summarize with Figure~\ref{coma}. 
The top left panel shows the distribution of all galaxies in the 2MASS Extended Source Catalog \citep{2000AJ....119.2498J} within a large region around the cluster.  Here, and with most other groups/clusters that will be discussed, the plots are in supergalactic coordinates because the entities tend to lie near the supergalactic equator.   Colors are coded by velocity with galaxies within 1000 \kms\ of the cluster mean in green and yellow.  The panels to the right and immediately below the first panel illustrate the velocities of galaxies within $3^{\circ}$ strips through the center of the cluster.  It can be seen that the cluster is being fed by filaments that extend in the SGB direction.  The cluster is abruptly bounded in SGL.   This characteristic is demonstrated most clearly in the lower right panel.  There is an edge to the cluster at $1.7^{\circ} = 3$~Mpc which is tentatively associated with the projected second turnaround radius $R_{2t}$.

The mass of the system can be estimated from application of the virial theorem with a sample of 367 galaxies in the 2MASS Extended Source Catalog within $1.7^{\circ}$ of the cluster center.  The dimension and velocity parameters are calculated using the bi-weight estimators discussed by \citet{1990AJ....100...32B}.  Here, and with all the groups, velocities are deprojected assuming $\alpha = 2.5$. The projected gravitational radius is $R_g = 2.15$~Mpc, the velocity dispersion is $\sigma_p = 954 \pm 50$~\kms, the virial mass is $M_{\rm v} = 1.8 \times 10^{15}~\Msun$, and the mass to K band light ratio is $127~\Msun / \Lsun$.   Here, the K band luminosity of $1.4 \times 10^{13}~\Lsun$ includes a correction of 60\% for lost light, a detail that will be discussed in Paper 2 of this series.

\noindent {\it The Virgo Cluster}, at a distance of 16.4 Mpc, is the nearest entity to attain Abell richness class zero (though not in the Abell catalog).   Because of its proximity, the Virgo Cluster has received particularly detailed attention \citep{2012ApJS..200....4F}. The interior of the cluster is composed of multiple components as manifested by the X-ray distribution \citep{1994Natur.368..828B} and is now somewhat resolved in three dimensions with accurate surface brightness fluctuation distances \citep{2007ApJ...655..144M, 2009ApJ...694..556B}.  It has long been known that the velocity structure of the cluster is complex. \citet{1985ESOC...20..181H} reports that the early type galaxies lie in close proximity to either M86-M87 or M49 and have a roughly gaussian distribution of velocities with dispersion $581 \pm 33$~\kms\ while the late type galaxies somewhat avoid the core and have a roughly boxcar distribution of velocities with dispersion $817 \pm 46$~\kms.  The most extreme blueshifted galaxies, surely legitimate members of the cluster, are distributed in an east-west band, evidence of a retained memory of a recent infall event.  \citet{1984ApJ...281...31T} discussed the infall of galaxies into the Virgo Cluster and pointed out that there will be an influx of spirals into the cluster over the next Hubble time that is comparable with the population that is already there.  A majority of these will arrive in the next 3 Gyr.

\begin{figure}[htbp]
%\begin{center}
%\includegraphics[scale=1]{virgo_extended_all6.ps}
%\special{psfile=virgo_extended_all6.ps hscale=108 vscale=108 voffset=-665 hoffset=-20 angle=0}
\includegraphics{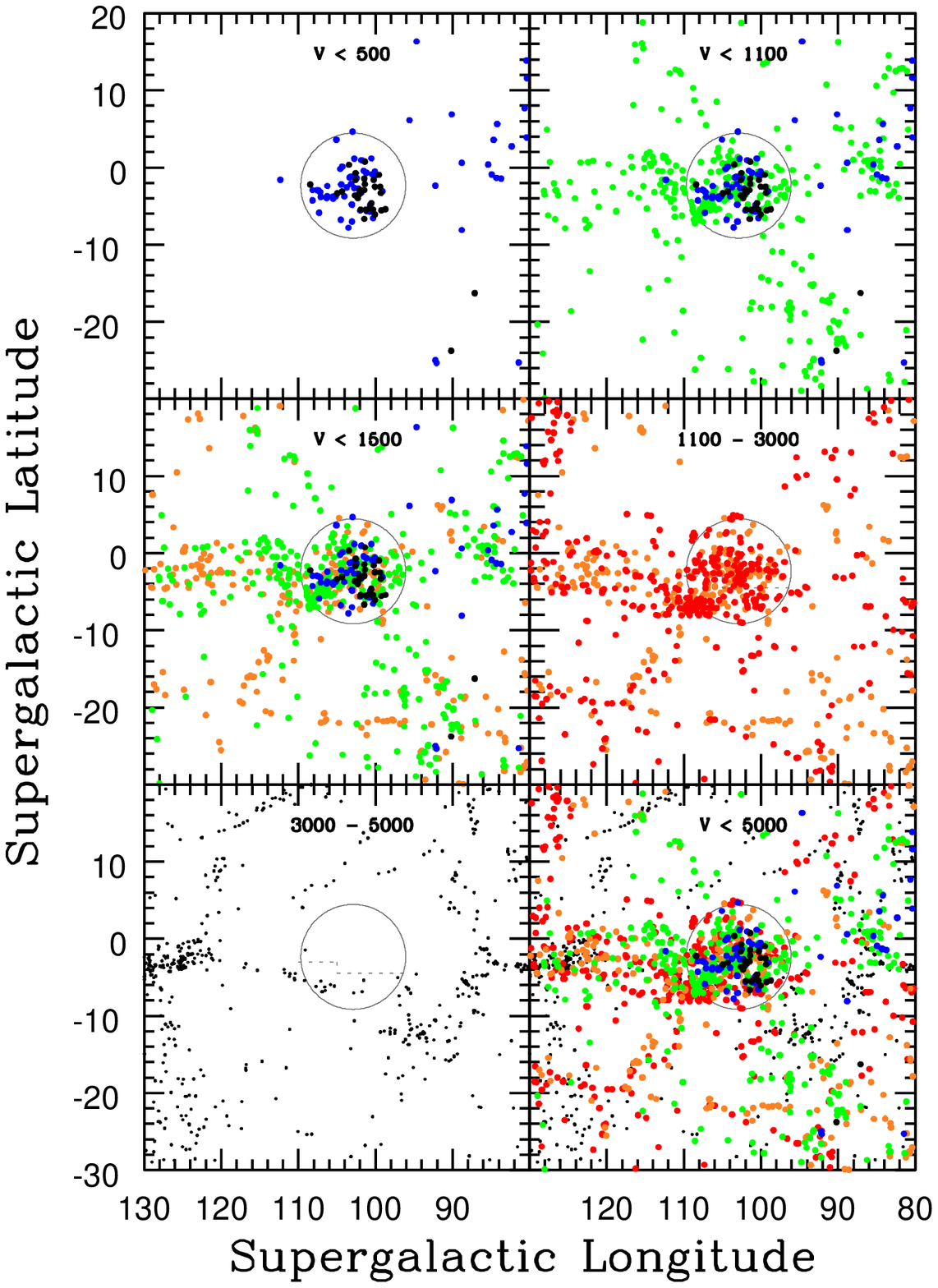}
\vspace{170mm}
\caption{Galaxies in the region of the Virgo Cluster with $B < 16$.  {\it Top left:} $V_{LS} < 0$~\kms\ in black and $0 < V_{LS} < 500$~\kms\ in blue.  {\it Top right:} Add $500 < V_{LS} < 1100$~\kms\ in green.  {\it Mid left:} Add $1100 < V_{LS} < 1500$~\kms\ in orange.  {\it Mid right:} Only $1100 < V_{LS} < 3000$~\kms\ with those above 1500~\kms\ in red.  {\it Bottom left:} Immediate background $3000 < V_{LS} < 5000$~\kms.  {\it Bottom right:} Entire sample with $V_{LS} < 5000$~\kms.  The circle in each panel is centered on M87, with $6.8^{\circ}$ radius.
}
\label{virgo}
%\end{center}
\end{figure}

There is an attempt with Figure~\ref{virgo} to illustrate the structure in the vicinity of the Virgo Cluster.  The sample is an extract from the Lyon Extragalactic Database (LEDA) of all galaxies in the plotted region with velocities less than 5000~\kms.  The sample was further restricted to $B < 16$ to avoid the very large contamination from spurious objects, particularly galactic stars, in a magnitude unlimited sample.  The few spurious objects remaining with the $B<16$ restriction were removed by hand.
Velocities, $V_{LS}$, are given in the rest frame of the Local Sheet \citep{2008ApJ...676..184T}.\footnote{The Local Sheet reference frame is similar to the several alternative variations of the Local Group reference frame.}

\onecolumn
\twocolumn

\onecolumn
\twocolumn

The plot in the top left panel shows the distribution of galaxies with velocities less than 500~\kms.  It is seen that most galaxies in this panel are strongly clustered.  The circle is centered on M87 and has a radius of $6.8^{\circ} = 2.0$~Mpc.  It will be argued below that this circle constitutes the projected radius of second turnaround, $R_{2t}$.  The black symbols identify galaxies with negative velocities.  These galaxies cluster tightly over a region within the circle from the core to the lower right.  The most negative high confidence velocity in the $B<16$ sample is $V_{LS} = -592$~\kms\ for IC 3492.  The most negative velocity in the unrestricted sample is $V_{LS} = -820$~\kms\ for VCC 846 \citep{1993A&AS...98..275B}.

The same galaxies are plotted on the top right panel and we add those with $500 - 1100$~\kms\ in green.  Together, these are the galaxies in the sample that have velocities less than roughly the mean for the cluster.  Within the $6.8^{\circ}$ circle a dense clump of objects is seen at SGL=107, SGB=$-$5 which is the sub-condensation around M49.  The $500 - 1100$~\kms\ window picks up a large number of objects near the supergalactic equator to the left of the $6.8^{\circ}$ circle, part of what is called the Virgo Southern Extension \citep{1961ApJS....6..213D}.  The other features that make an appearance in this panel are parts of what are called the `Ursa Major Cloud' (center right), the `Leo Spur' (lower right), and the `Virgo-Libra Cloud' (upper left) in the Nearby Galaxies (NBG) Atlas \citep{1987nga..book.....T}.  

The same objects are shown in the mid-left panel, now with the addition of galaxies in the range $1100 - 1500$~\kms.  The purpose for drawing attention to this small incremental step in velocities is to illustrate that galaxies in the Southern Extension have modest velocity dispersions with respect to their immediate neighbors.  The green and orange points separate in this region.  The structure that has now appeared to the lower left is the `Crater Cloud' in the Nearby Galaxies Atlas.

The mid-right panel shows the distribution of galaxies with velocities in the range from roughly the mean to the high velocity extreme of the cluster.  The main new feature to note is the appearance of de Vaucouleurs' Virgo W Cluster at  SGL=108, SGB=$-$7.  This significant entity is known to be in the background at twice the distance of the Virgo Cluster.  Its mean velocity is $V_{LS} = 2261$~\kms\ and distance is 34 Mpc.  Virgo W lies at radii $6^{\circ} - 9^{\circ}$ from M87 so it surely contaminates the quasi-virialized region of the Virgo Cluster.  Other background contaminants are known.  Both Cepheid and SNIa distances for NGC~4639 and 5 surface brightness fluctuation distances places the Virgo W$^{\prime}$ group at 23.7 Mpc, 7.3 Mpc directly behind the Virgo Cluster.  Then the M Cloud \citep{1984ApJ...282...19F} is determined to lie at 35 Mpc, similar to the distance for Virgo W.  Details on current distance measurements for these background entities are given by  \citet{2013AJ....146...86T} and \citet{2014ApJ...782....4K}.   In Table 2 of Tully et al., W is group icnt = 56 (NBG catalog 11-24). W$^{\prime}$ is group icnt = 5 (NBG 11-5), and M is group icnt = 177 (NBG 13+12).   In the mid-right panel of Fig.~\ref{virgo} there is a horizontal dashed line with a dog-leg. Almost all the known background contaminants within the $6.8^{\circ}$ circle lie  below this line.  The Virgo Cluster suffers from contamination that cannot be distinguished with velocity information but known cases lie over a restricted part of the cluster.

The lower left panel shows the distribution of background galaxies in the range 3000--5000~\kms.  It is seen that the cluster is relatively clear of contaminants from the near background and what exists is confined to a band at SGB~$\sim -6$, below the dogleg dashed line introduced in the previous panel.  These objects can be distinguished with redshifts.  The void behind Virgo extends to the Great Wall at 6000~\kms (and probably accounts for an apparent peculiar velocity of the cluster toward us).  Finally, in the lower right panel all the information from the previous panels is combined.

\begin{figure}[htbp]
\begin{center}
\includegraphics[scale=0.38]{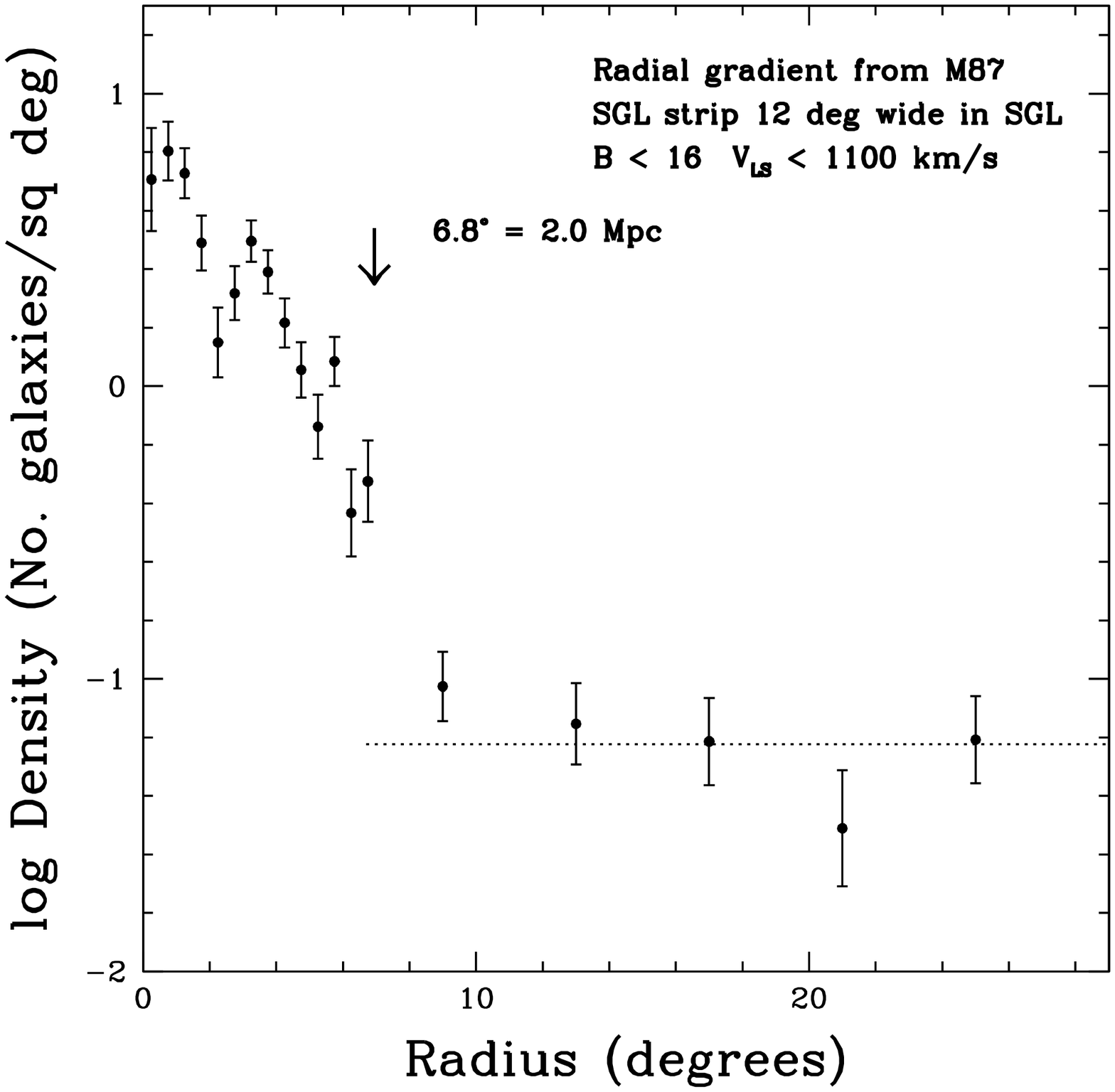}
\caption{Surface density of galaxies ($B < 16$ and $V_{LS} < 1100$ km/s) as a function of distance from M87 in a strip $12^{\circ}$ wide centered on M87 in SGL and running from $-$29 to +25 in SGB.  A discontinuity is seen at $6.8^{\circ} \pm 0.4$.
}
\label{virgo_rad}
\end{center}
\end{figure}

Conveniently, as with Coma, the current infall into the Virgo Cluster is along a cardinal axis in supergalactic coordinates.  The constraints on the dimension of the cluster is investigated by considering the distribution of galaxies along a strip $12^{\circ}$ wide in SGL centered on M87 and extended in SGB (with $V_{LS} < 1100$~\kms\ to minimize contamination).  The results are shown in Figure~\ref{virgo_rad}.  It is seen that there is an abrupt edge to the cluster at $6.8^{\circ} \pm0.4$ which is inferred to correspond to the projected radius of second turnaround, $R_{2t}$.

At larger radii, the Virgo Southern Extension at more positive values of SGL is well populated and the velocity dispersion is large if one does not spatially discriminate.  However, as mentioned in connection with the mid-left panel of Fig.~\ref{virgo}, velocity--spatial separation can be discerned in the Southern Extension.  Models have been developed that describe the infall toward Virgo \citep{1984ApJ...281...31T, 1999A&A...351..827E, 2014ApJ...782....4K}.  In some cases distance measurements are accurate enough to confirm the infall prediction that galaxies in the Southern Extension blueshifted with respect to the cluster mean are more distant than the cluster and galaxies redshifted with respect to the cluster are nearer.  The mass of the cluster can be estimated from a Numerical Action orbit reconstruction of the infall to be $8 \times 10^{14}~\Msun$ \citep{2005ApJ...635L.113M}.  A consistent value is found from a spherical infall model  \citep{2014ApJ...782....4K}. Numerical Action models suggest that infalling galaxies have orbital angular momentum acquired from large scale tides, arriving in the cluster with radial velocities of 1500~\kms\ and tangential components of $\sim 100$~\kms\ \citep{1995ApJ...454...15S}.

The essential point for the current discussion is that the infall and quasi-virialized domains of the Virgo Cluster can be distinguished, characterized by a projected second turnaround radius $R_{2t} = 6.8^{\circ} \pm 0.4$.  If, now, a virial analysis is carried out, with elimination of known or suspected contaminants, one finds the parameters $<V_{LS}> = 1083$~\kms, $\sigma_p = 732\pm45$~\kms, and $M_{\rm v} = 7 \times 10^{14}~\Msun$. The virial and infall model mass estimates are in good agreement.

\subsection{Two Intermediate Mass Evolved Groups}

\begin{figure}[htbp]
\begin{center}
\includegraphics[scale=0.38]{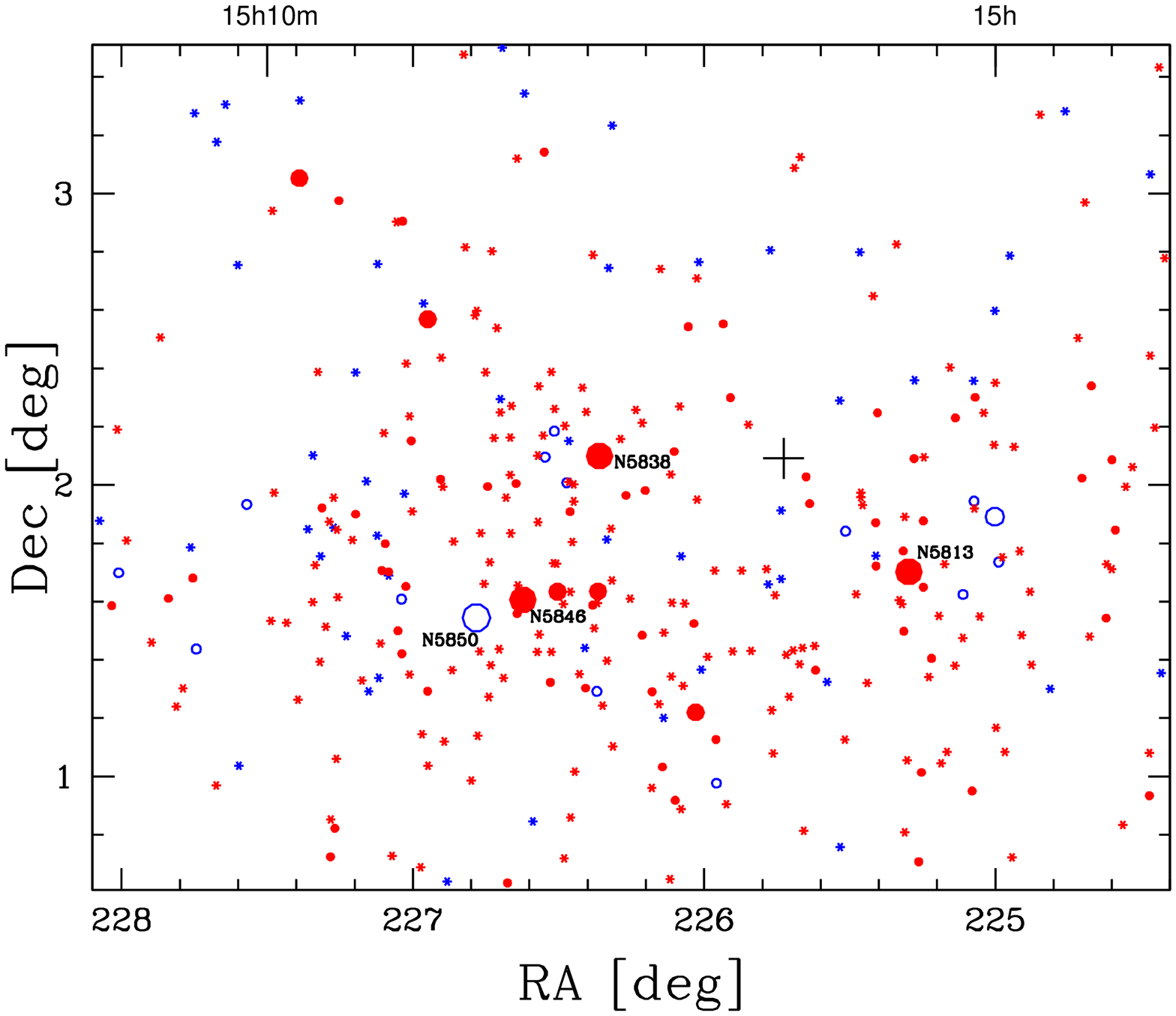}
\includegraphics[scale=0.4]{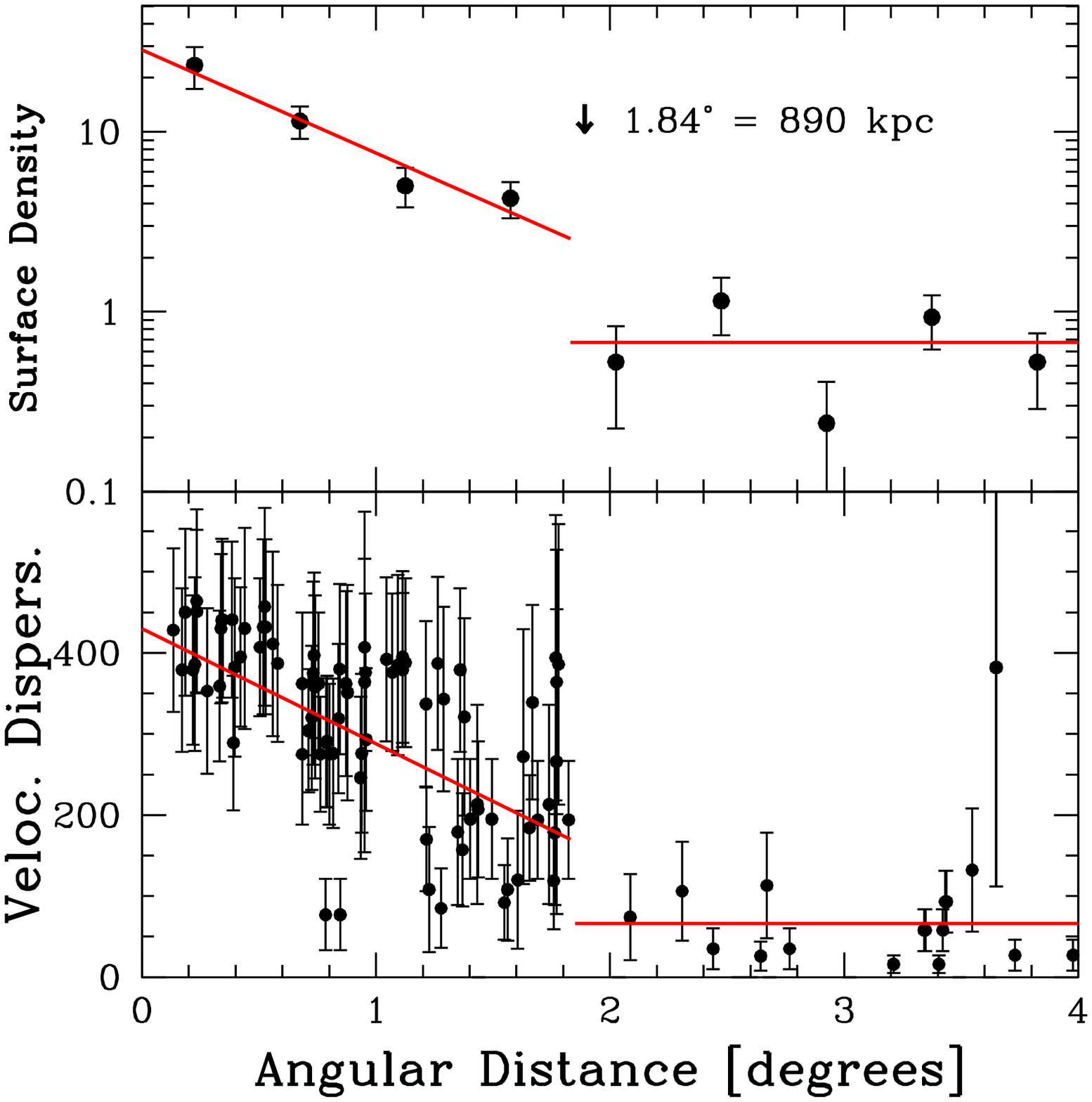}
\caption{{it Top:} Distribution of galaxies in the vicinity of the NGC 5846 Group.  Red:  E/S0 galaxies;  open blue: spiral/irregular galaxies.  Stars: unknown velocities.  {\it Middle:} Run of surface density with radius.  {\it Bottom:} Run of velocity dispersion with radius.  Each data point represents the dispersion in the local proximity to each galaxy in the sample as described by Mahdavi et al. (2005).  Beyond $2^{\circ}$, the {\it local} velocity dispersion is low.
A discontinuity at $1.84^{\circ} = 890$ kpc is seen in each of the two bottom panels. }
\label{n5846}
\end{center}
\end{figure}

\noindent {\it The NGC 5846 Group}, at an assumed distance of 27.5 Mpc, is dominated by two giant ellipticals.  Most known members are early types.  This group was selected for intensive study as part of the CFHT wide field imaging program \citep{2005AJ....130.1502M} because it is relatively rich (250 galaxies brighter than $M_R=-11$) and lies within a filament that projects near the plane of the sky in front of a substantial void. The following information is drawn from Mahdavi et al. with adjustments due to a revised distance. In the three panels of Figure~\ref{n5846} one sees the distribution of known galaxies at roughly the distance of the group, the density distribution as a function of radius from the group center, and the velocity dispersion as a function of this radius.  There are clear density and velocity dispersion discontinuities at $R_{2t}=890$ kpc.  The group velocity dispersion of $\sigma_p=322\pm35$ \kms\ is well established by 83 measurements.  The inferred virial mass for the group is $8 \times 10^{13}~\Msun$.  

\noindent {\it The NGC 1407 Group}, at 26.3 Mpc, almost qualifies as a `fossil group' \citep{1994Natur.369..462P} with a difference in magnitude between the dominant elliptical and the second brightest member of 1.4 mag (a difference of 2 mag is required to meet the qualification of a fossil group).  Extended X-ray emission is seen from a hot intra-group plasma but it is faint.  The group was selected for study during the CFHT wide field imaging survey as a dense accumulation of overwhelmingly early type galaxies nicely isolated from foreground and immediate background confusion \citep{2006MNRAS.369.1375T}.  Some 240 galaxies are associated with the group brighter than $M_R = -11$.  Velocities are now available for 69 of these, including material from \citet{2009AJ....137.4956R} and two velocities given below.  Figure~\ref{n1407} illustrates the run of surface density and, particularly interesting, of velocity with radius from the dominate central galaxy.  The group has received considerable attention because of the large relative blueshift of the second brightest galaxy, NGC 1400 \citep{1993ApJ...403...37G, 1994A&A...283..722Q}.  With our recent spectroscopy with Subaru Telescope we have found two dwarfs with relative blueshifts that are almost as extreme: \citet{2006MNRAS.369.1375T}  ID 79 with $V_{helio}=656$~\kms\ and ID 96 with $V_{helio}=603$~\kms.  The likely explanation is the recent infall of NGC 1400 and a small entourage.  NGC 1400 must be near the potential minimum with a velocity vector fortuitously pointing almost at us.  Inferred properties for the group are $R_{2t} \sim 900$~kpc (uncertain because at the limit of the Trentham et al. survey), $\sigma_p = 365\pm44$~\kms, $M_{\rm v} = 6 \times 10^{13}~\Msun$, and $M_{\rm v} / L_R = 240~\Msun/\Lsun$. 

\begin{figure}[htbp]
\begin{center}
\includegraphics[scale=0.4]{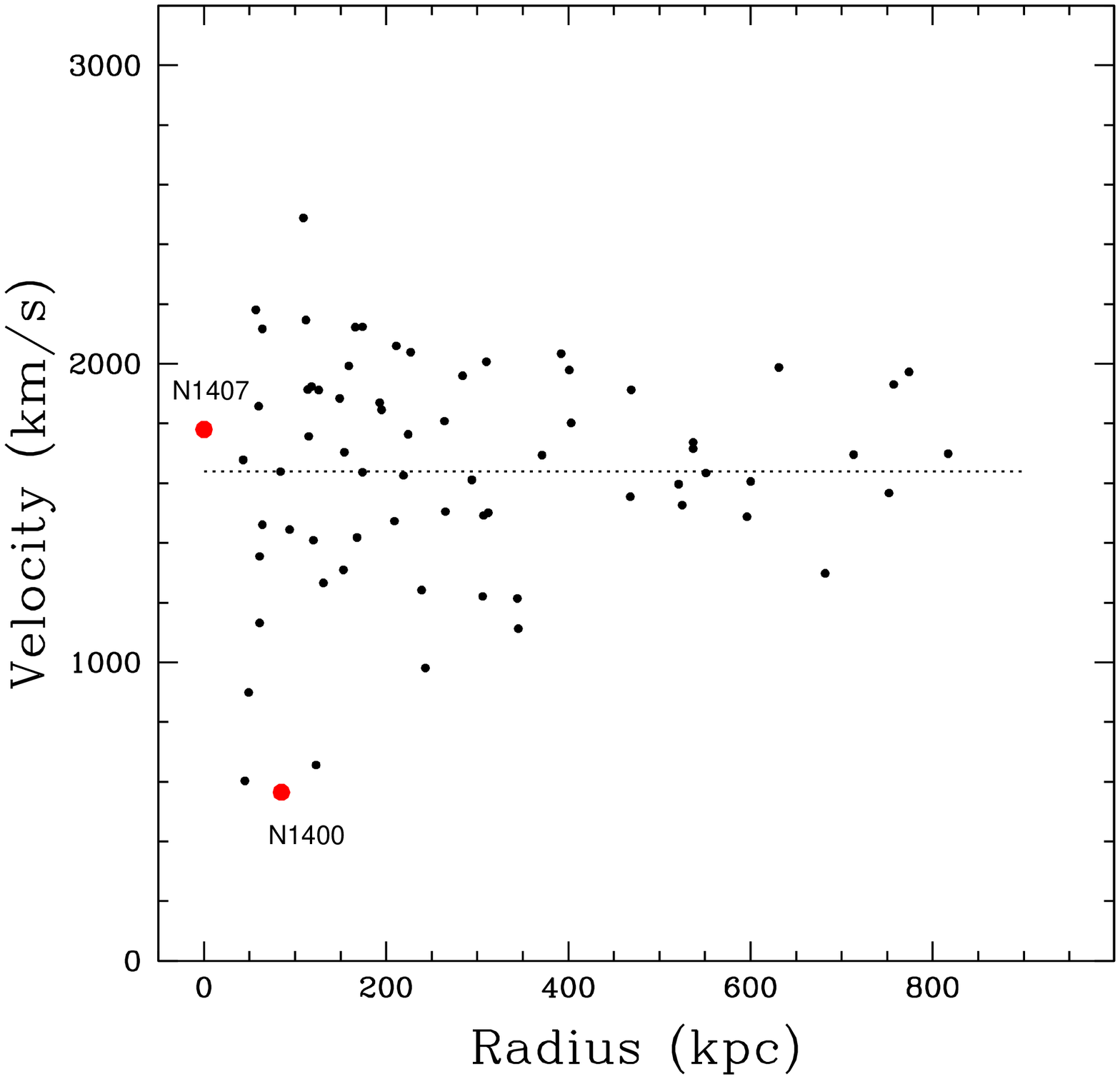}
\includegraphics[scale=0.4]{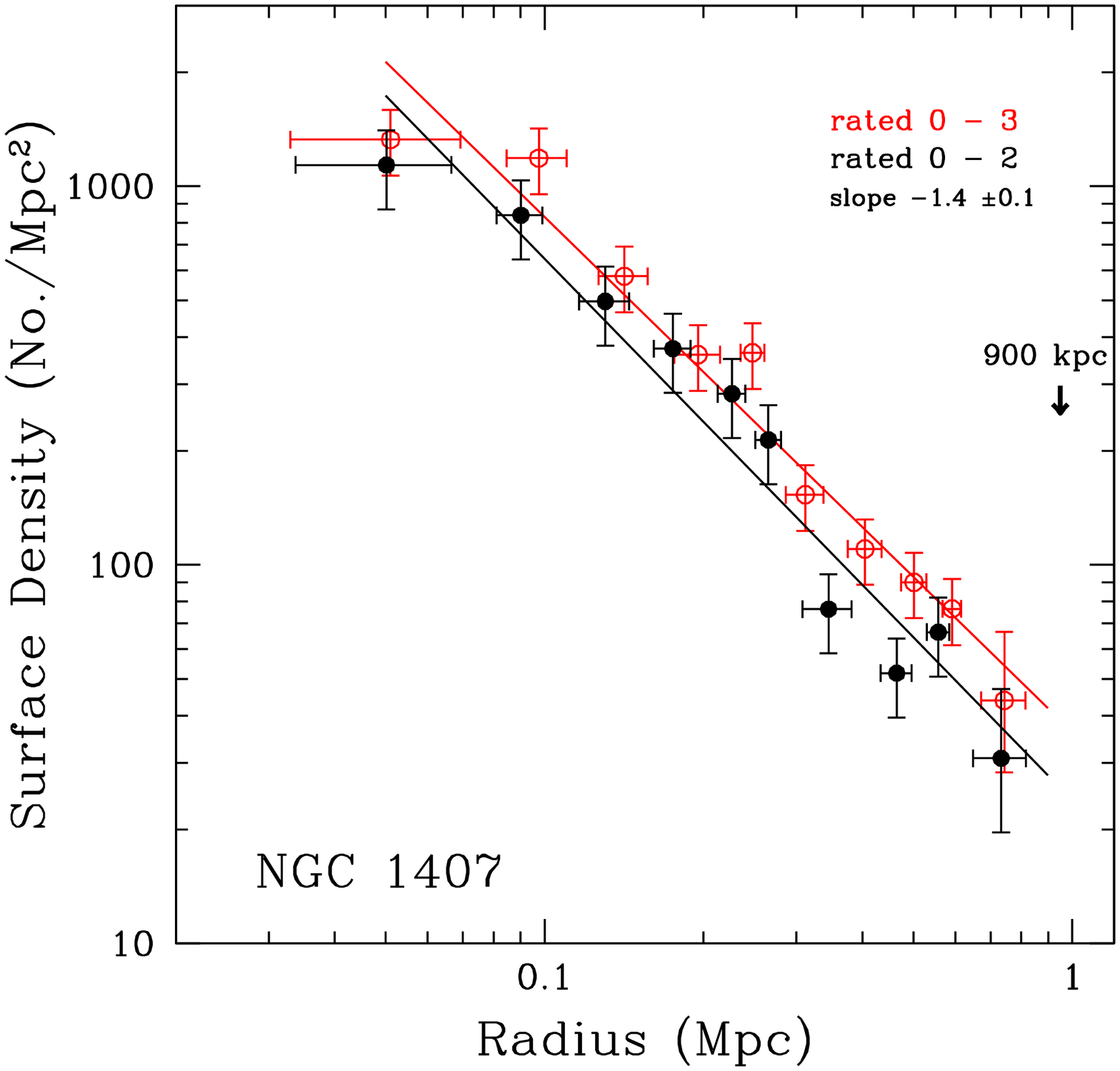}
\caption{{it Top:} Velocities as a function of distance from NGC 1407.  The extreme blueshifted velocity of the second brightest galaxy in the group, NGC 1400, is shared by two dwarfs.
{\it Bottom:} Run of the surface density of galaxies with distance from NGC 1407. Ratings 0$-$2 are high probability members; 3 are uncertain.
}
\label{n1407}
\end{center}
\end{figure}

\begin{figure}[htbp]
\begin{center}
\includegraphics[scale=0.47]{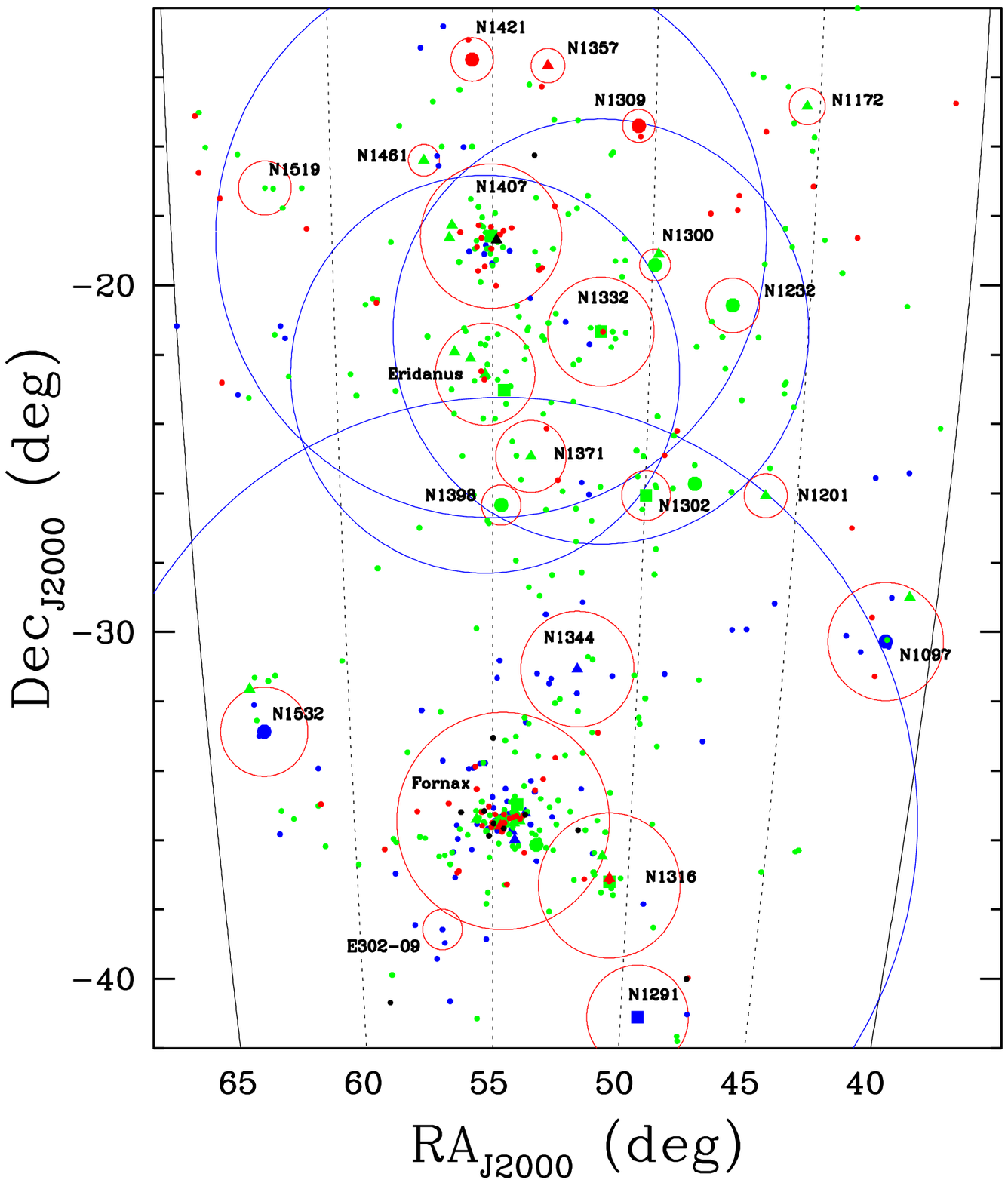}
\caption{All galaxies with $2^h40^m < RA_J < 4^h20^m$, $-42 < DE_J < -12$, $V_h < 2400$~\kms.  Black: 500-800~\kms; blue: 800-1300~\kms; green: 1300-1900~\kms; red: 1900-2400~\kms.  Red/blue rings: second and first turnaround radii around principal groups.
}
\label{eri_for}
\end{center}
\end{figure}

\begin{figure}[htbp]
\begin{center}
\includegraphics[scale=0.4]{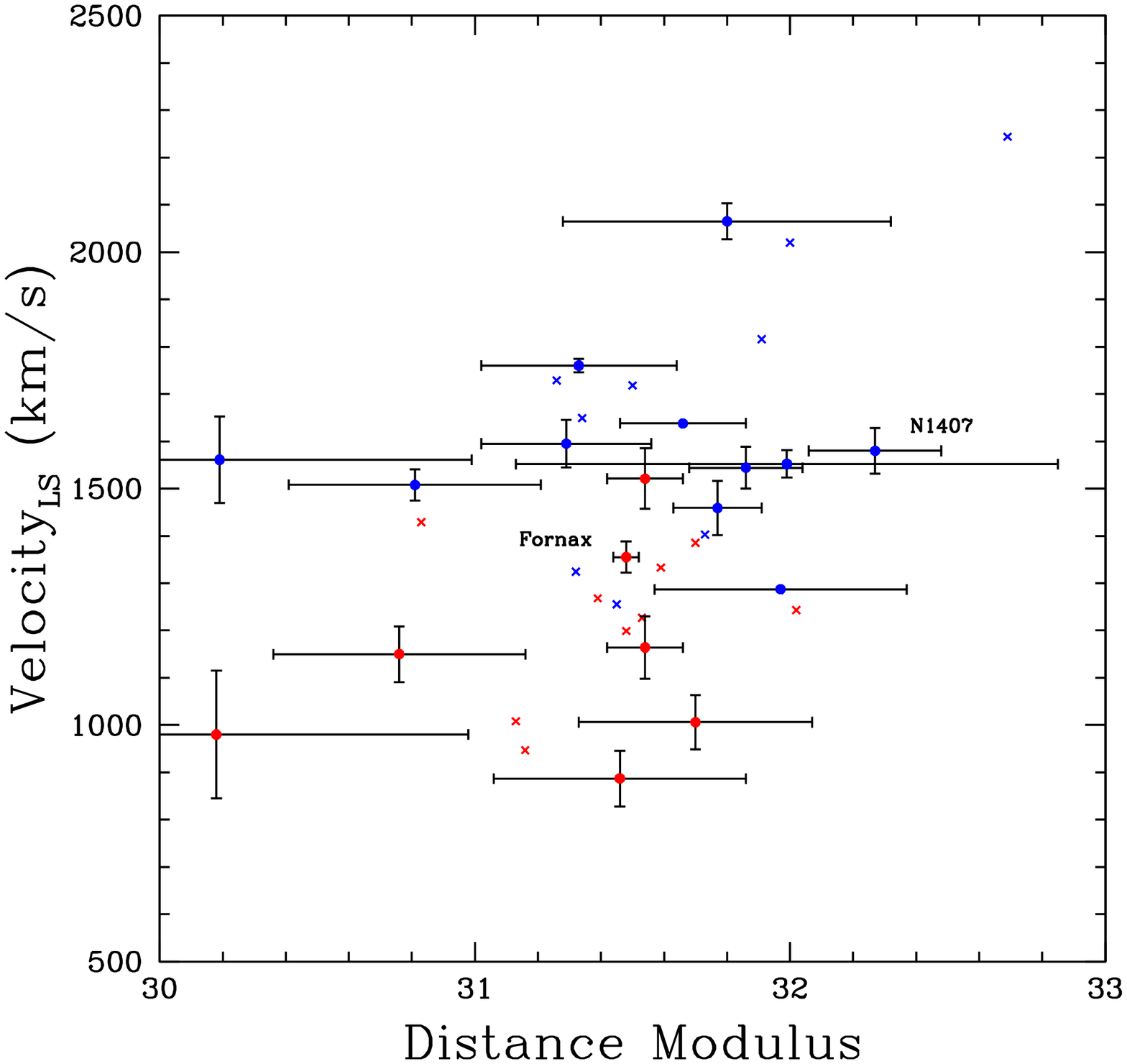}
\caption{
Distance moduli vs. Local Sheet velocities. Points with error bars: principal groups; crosses: individual galaxies in the field.  Red: nearest Fornax; blue nearest NGC 1407.
}
\label{eri_for-2}
\end{center}
\end{figure}

The NGC 1407 Group is part of the Eridanus Cloud, itself in close proximity to the Fornax Cluster.  The dynamical status of the Eridanus region has been studied by \citet{2006MNRAS.369.1351B} who reasonably conclude that the structure is bound. In fact, the entire region from Fornax Cluster to the NGC 1407 Group is probably destined to collapse within another Hubble time.  Figure~\ref{eri_for} is a plot of the projected distribution of galaxies in the region.  Figure~\ref{eri_for-2} compares velocities and distances. The point to the lower left in Fig.~\ref{eri_for-2} stands for the NGC 1097 Group which is probably to the foreground.  Otherwise, there is little hint of a correlation of velocity with distance.  Unfortunately, only the distance to the Fornax Cluster \citep{2009ApJ...694..556B} is known with the sort of precision required for a proper dynamic analysis of the region. 

In the region, in addition to NGC 1407 and Fornax, the Eridanus (NGC 1395) and NGC 1332 groups are significant.  The Fornax Cluster, at 18.8 Mpc, with $\sigma_p = 311\pm44$~\kms, and $M_{\rm v} = 9 \times 10^{13}~\Msun$, has indications of an infall zone at radii less than 5 Mpc (Fig.~\ref{eri_for}).  The Eridanus and NGC 1332 groups have velocity dispersions 228 and 205 \kms, respectively, and masses in the range $1-2 \times 10^{13}~\Msun$. Globally the environment includes $3-4 \times 10^{14}~\Msun$.  Fig.~\ref{eri_for} includes estimates of the radii of first turnaround, $R_{1t}$, around the various groups (justified in a later section).  These circles correspond to spheres that are strongly overlapping in three-dimensions, evidence that the entire region is bound and collapsing.  The NGC 1407 Group may be a (near) fossil group today but it is destined to de--fossil.

\subsection{A Group with Mixed Morphology}

\begin{figure}[htbp]
\begin{center}
\includegraphics[scale=0.44]{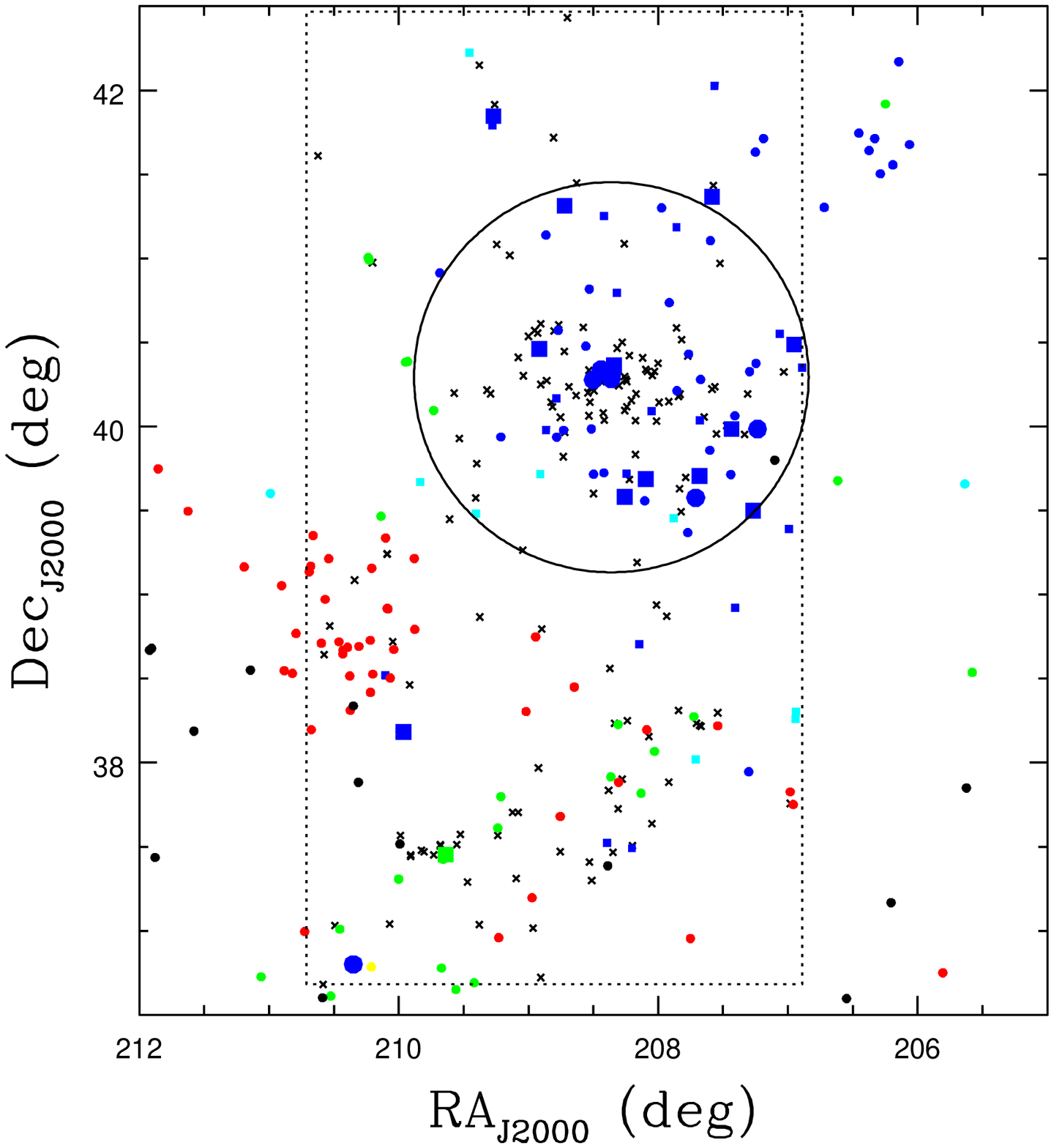}
\includegraphics[scale=0.4]{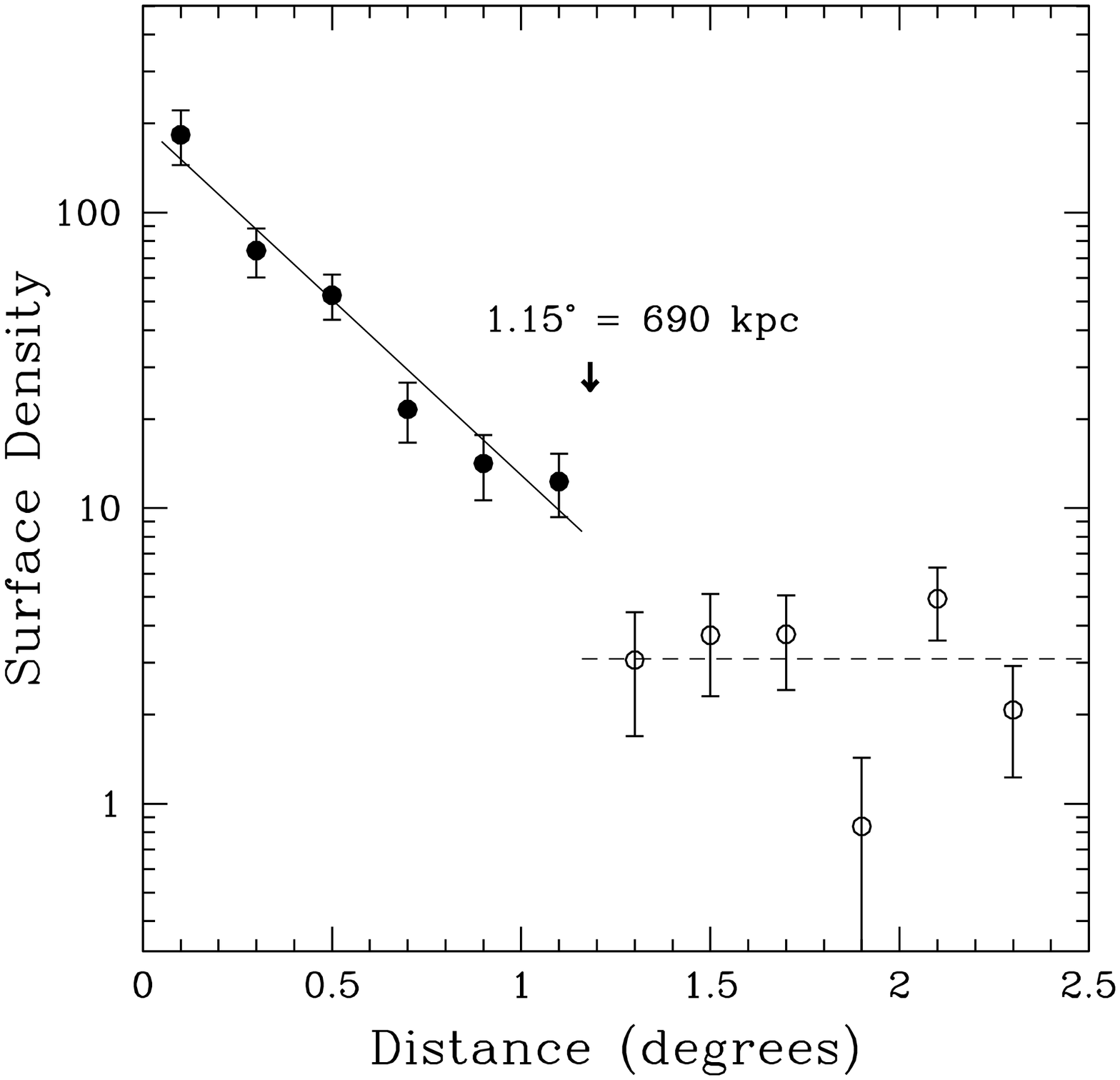}
\caption{{\it Top:} Distribution of galaxies in the vicinity of the NGC 5353/54 Group.  Dark Blue: galaxies with velocities associated with the group.  Cyan: foreground; green and red: background.  Black crosses: unknown velocities.  Dashed box: region of CFHT Megacam survey. {\it Bottom:} Run of surface density with radius.  A discontinuity at $1.15^{\circ} = 690$ kpc is seen in the bottom panel.  A circle of this radius is superimposed on the top panel. }
\label{n5371}
\end{center}
\end{figure}

\noindent {\it The NGC 5353/4 Group} was studied during the CFHT campaign and discussed by   \citet{2008AJ....135.1488T}.  Again, the group was selected in part because of its apparent isolation.  The panels of Figure~\ref{n5371} illustrate the situation.  The group contains 126 galaxies brighter than $M_R=-11$, 53 with known velocities.  A discontinuity identifies the second turnaround at $R_{2t}=690$ kpc if the distance is 34.7 Mpc.  The velocity dispersion is $209\pm29$ \kms\ and the virial mass is $3 \times 10^{13}~\Msun$.

A particular interest with this group is the mixed morphology.  A core of early type galaxies has apparently recently received an influx of late type systems.   There is an unusually large number of transition dwarfs; small galaxies with dE morphologies but with emission line or A star absorption features.

\subsection{A Biggest of the Small}

\begin{figure}[htbp]
\begin{center}
\includegraphics[scale=0.53]{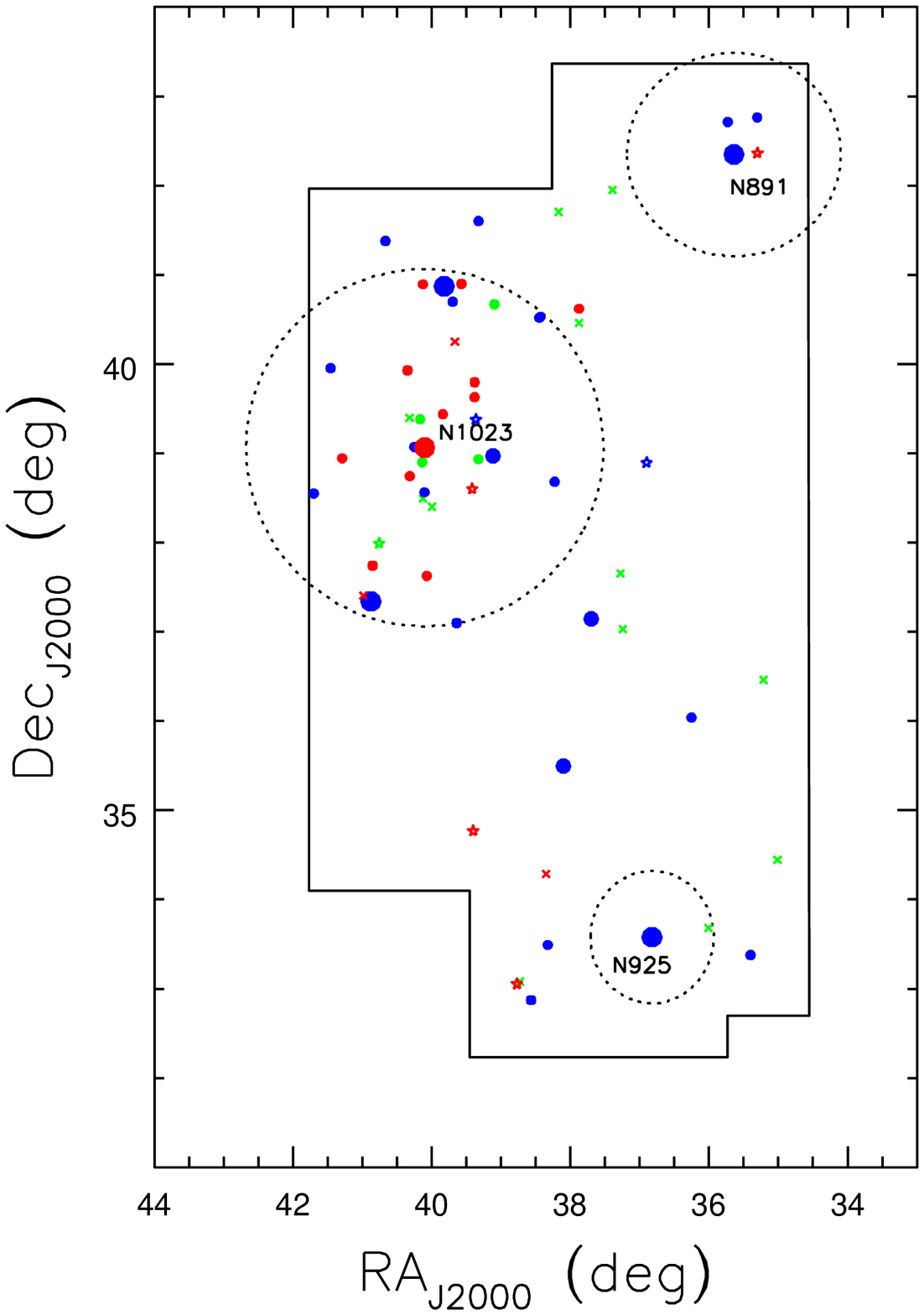}
\caption{Distribution of galaxies in the vicinity of the NGC 1023 Group.  The irregular box is the CFHT survey region.  Red: early types; blue: late types; green: transition or uncertain types. }
\label{n1023}
\end{center}
\end{figure}

\begin{figure}[htbp]
\begin{center}
\includegraphics[scale=0.41]{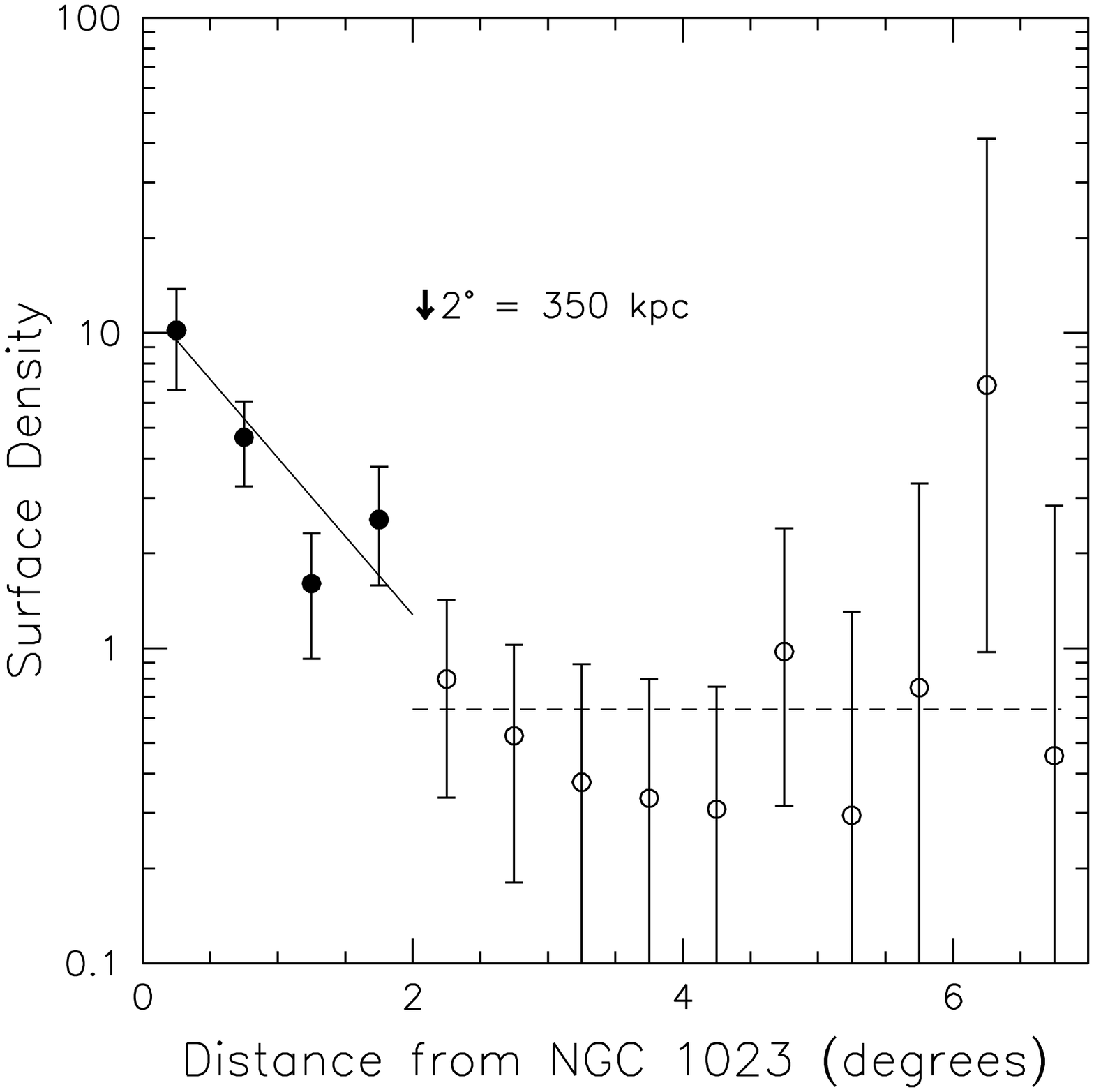}
\includegraphics[scale=0.4]{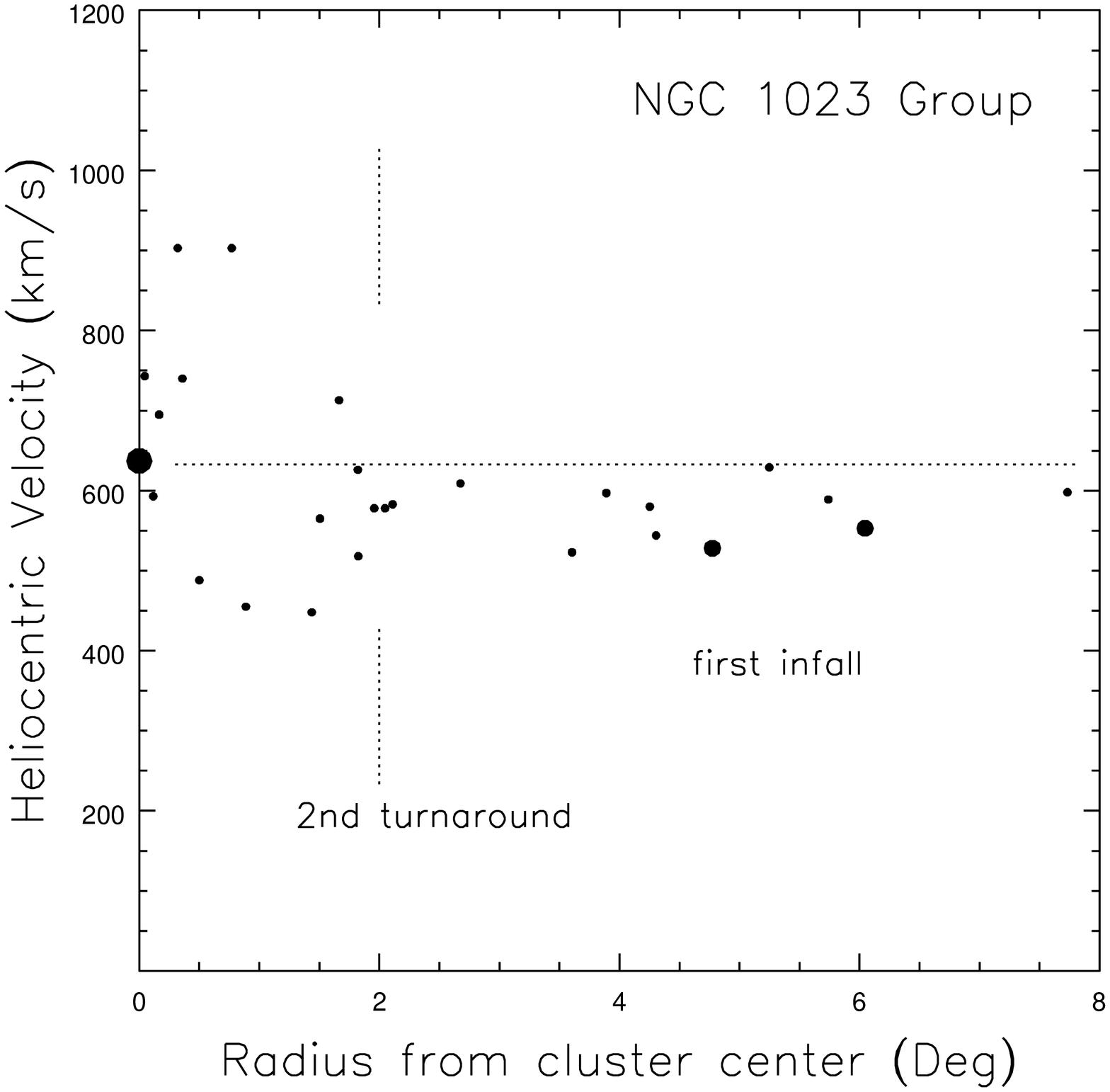}
\caption{{\it Top:} 
Run of surface density with radius.  {\it Bottom:}  Velocities as a function of radius.  A discontinuity in density and transition to low velocity dispersion occurs at $2^{\circ} = 350$ kpc.  A circle of this radius is drawn around NGC 1023 in Fig.~\ref{n1023}. }
\label{n1023-2}
\end{center}
\end{figure}

\noindent {\it The NGC 1023 Group} was imaged at CFHT over an extended region \citep{2009MNRAS.398..722T}.  See the distribution of galaxies on the sky with Figure~\ref{n1023} and the runs of surface density and velocity in Figure~\ref{n1023-2}. The statistics are poor with such a small group but $R_{2t}$ around NGC 1023 can be identified at 350 kpc.  Within this radius there are 41 suspected members, 18 with known velocities.  At a distance of 10.1 Mpc, $R_{2t}=350$ kpc, $\sigma_p=133\pm31$ \kms, and the virial mass is $5 \times 10^{12}~\Msun$.  The survey region was extensive enough to document the "Dressler effect"  \citep{1980ApJ...236..351D}; the prominence of early types near the core and late types at large radii.  Specifically, the transition can be associated with the radius $R_{2t}$.

\subsection{The Nearest Groups}

Groups with mass below $5 \times 10^{12}~\Msun$ are so poorly populated that they are hard to study at large distances, yet it is theoretically anticipated and observed in our own vicinity that such groups are common.  Thanks to observations with HST, there is now a good understanding of the detailed distribution of galaxies within 4 Mpc, with distance measures with uncertainties less than 200 kpc.  For most of these systems there are velocities accurate to $\sim 5$~\kms\ from HI observations with radio telescopes.  Within 4 Mpc at unobscured Galactic latitudes there are four well known complexes of galaxies: those of the Local Group, around Centaurus A, M81, and in the constellation of Sculptor.  In addition, there are three associations containing only dwarf galaxies.  These associations will be discussed in a later section.

\begin{figure}[htbp]
\begin{center}
\includegraphics[scale=0.4]{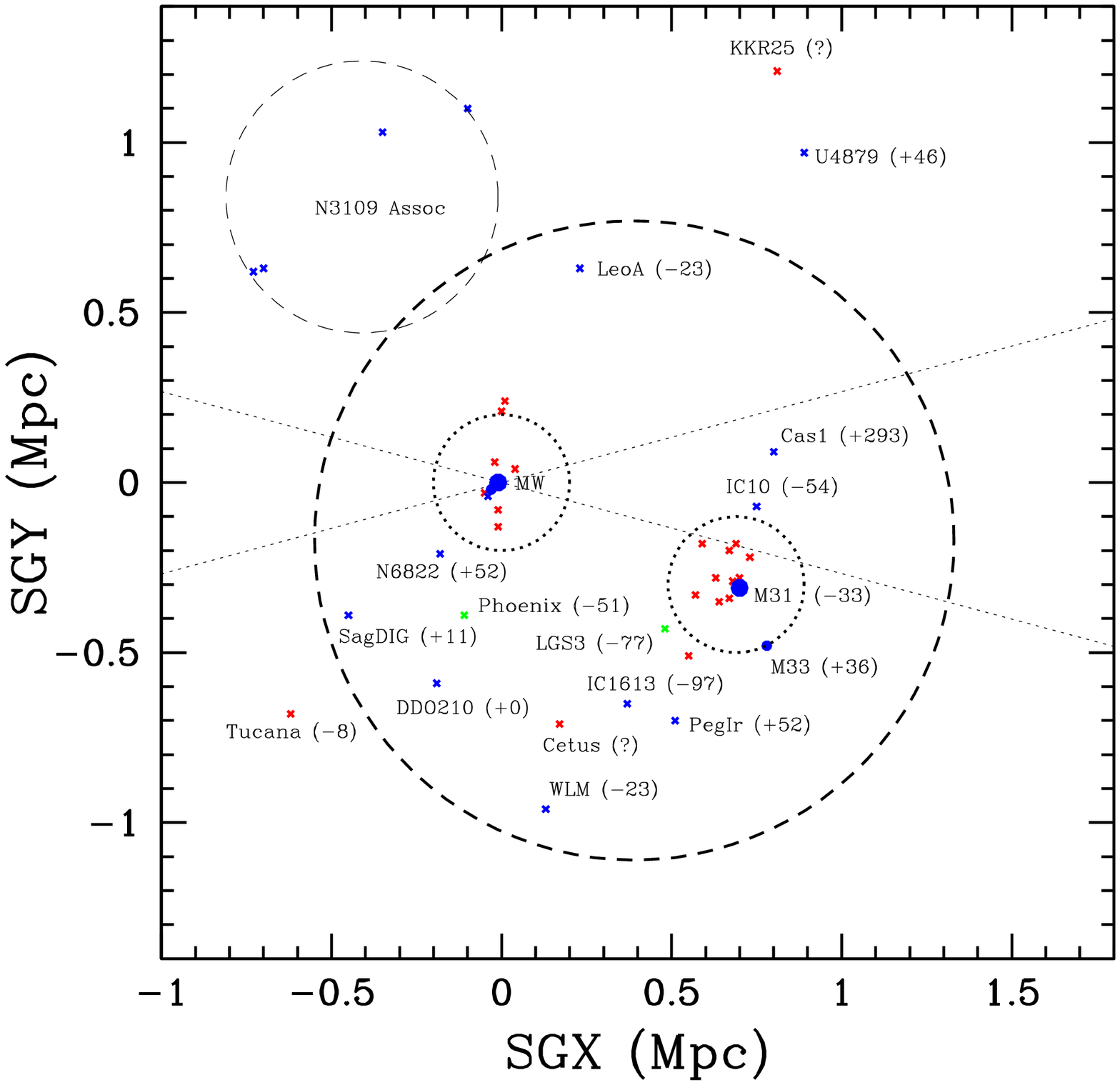}
\includegraphics[scale=0.4]{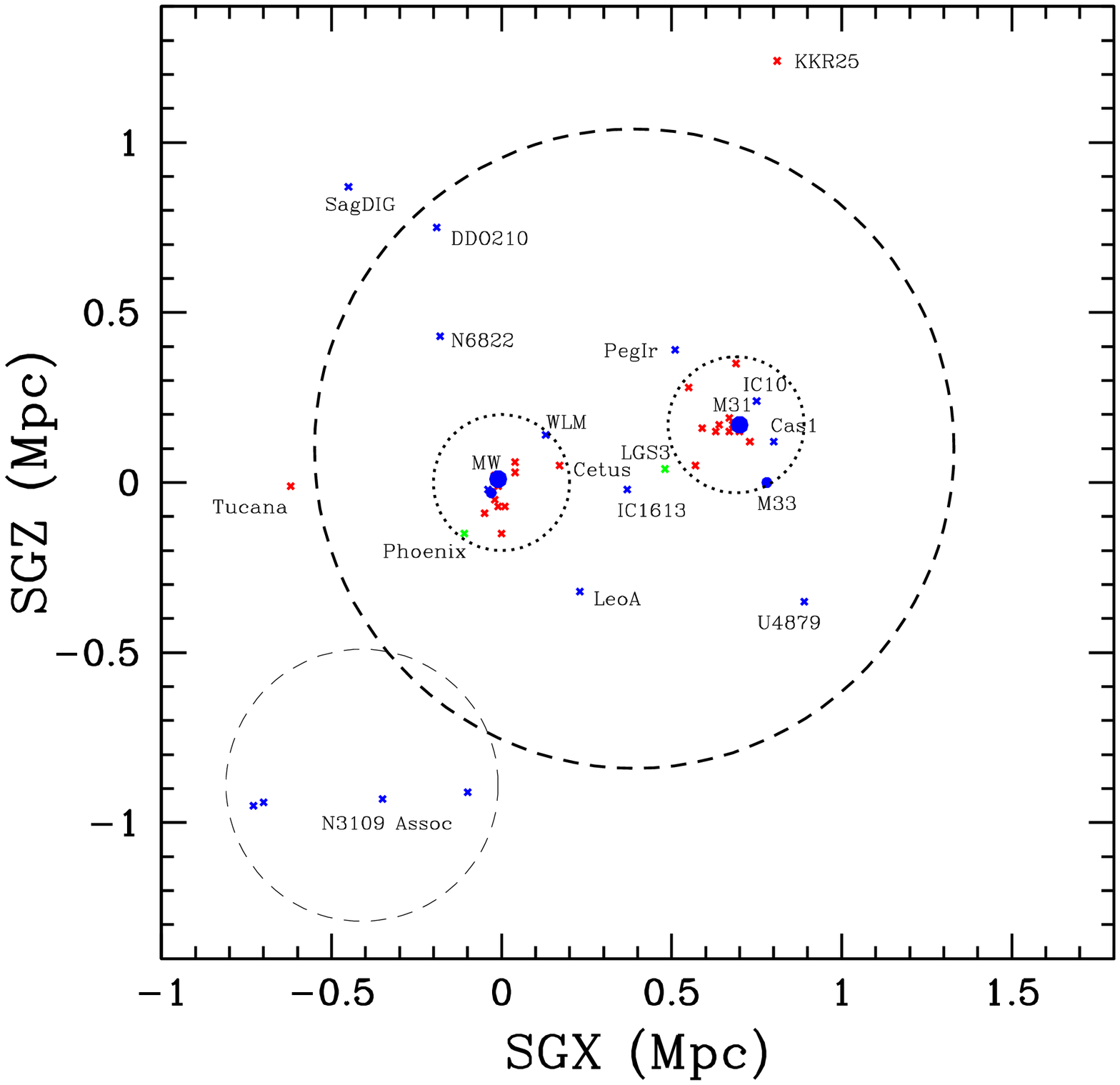}
\caption{Two projections of the nearest galaxies in supergalactic coordinates.  Red: early types; green: transition types; blue: late types.  The inner and outer circles approximate the radii of second and first turnaround, respectively.  The circle near the edge of the fields encloses the distinct NGC 3109 Association.  Galactic latitudes $\vert b \vert < 15^{\circ}$ lie within the wedges centered on SGY = 0.}
\label{1Mpc}
\end{center}
\end{figure}

\noindent {\it The Milky Way and M31} have colloquially been said to be members of the Local Group but it is now understood that these two large galaxies lie in separate halos within a common infall zone.  Figure~\ref{1Mpc} illustrates the three-dimensional distribution of the nearest galaxies.    The masses of the two dominant halos can be probed by the motions of gas, stars, and satellite galaxies with greater detail than in more distant environments.  Recent estimate of the virial mass of the Milky Way range from $1.0 \times 10^{12}~\Msun$ \citep{2008ApJ...684.1143X} to $1.9 \times 10^{12}~\Msun$ \citep{1999MNRAS.310..645W}.   Recent estimates for M31 range from $0.7 \times 10^{12}~\Msun$ \citep{2006MNRAS.366..996G} to $1.6 \times 10^{12}~\Msun$ \citep{2002ApJ...573..597K}, a slightly lower range than for the Milky Way even though M31 is 40--50\% more luminous.  Recent observations of the proper motion of M31 help to restrict the range of plausible masses.  The constraints have been reviewed by \citet{2012ApJ...753....8V} who find virial masses of $1.5\pm0.4 \times 10^{12} \Msun$ for M31 and $1.6 \times 10^{12} \Msun$ for the Milky Way with more ambiguous uncertainty.

\begin{figure}[htbp]
\begin{center}
\includegraphics[scale=0.61]{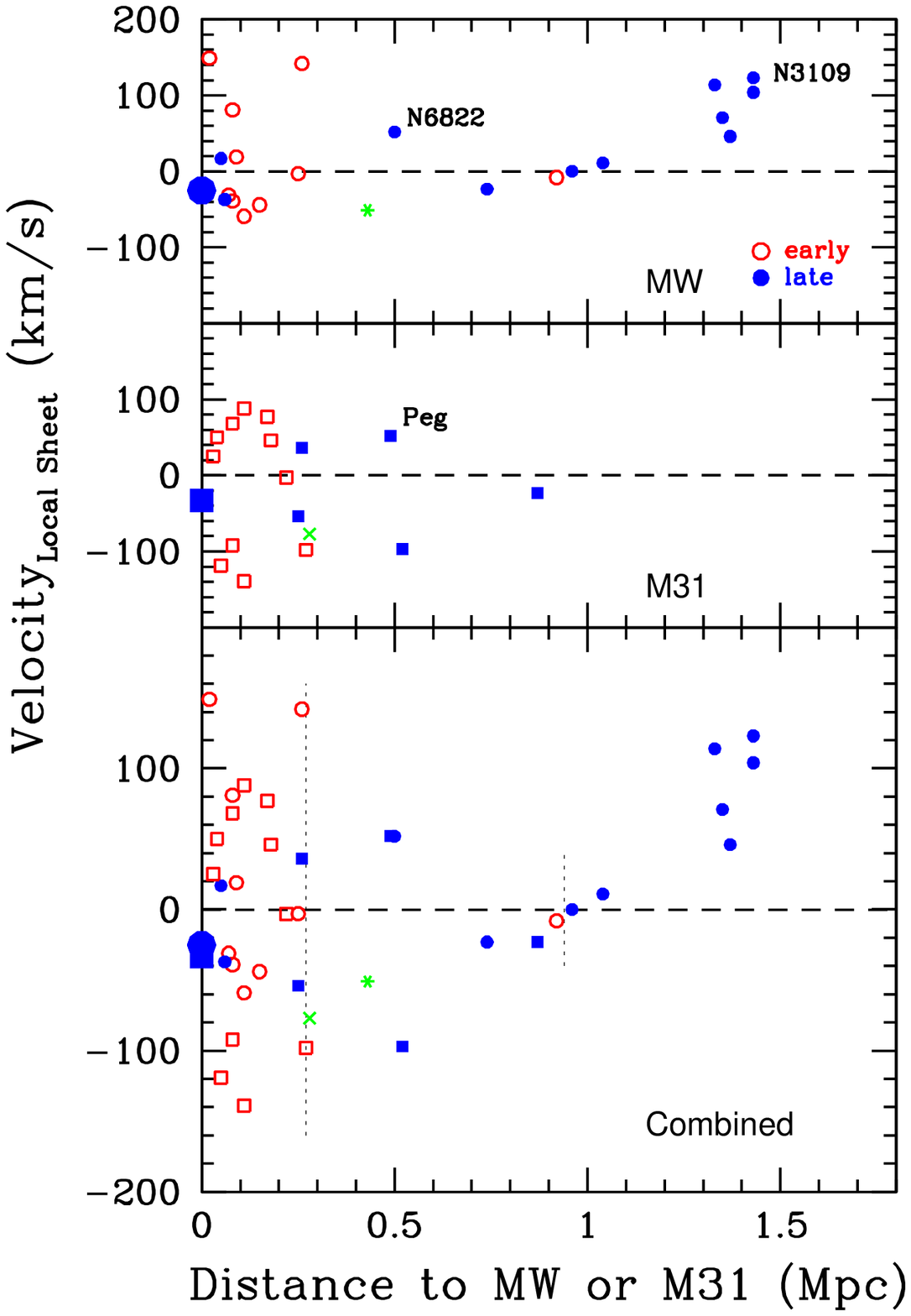}
\caption{Velocities of the nearest galaxies vs. the distance from the Milky Way (top) or M31 (middle), whichever is nearer.  The combined data is shown in the bottom panel.  Open red: early types; green star or cross: transition types; filled blue: late types.  Galaxies within 280 kpc of one of the two dominant galaxies have large velocity dispersions and are typically early types.  Galaxies at larger radii are predominantly late types and have the velocity characteristics of first infall within $\sim 940$ kpc and cosmic expansion outside this radius.} 
\label{MW-M31}
\end{center}
\end{figure}

The Milky Way and M31 second turnaround halos are discrete but the two galaxies lie within a common infall zone approximated by the larger circles in Fig.~\ref{1Mpc}.   The infall is clearly seen in Figure~\ref{MW-M31}.  Velocities of galaxies in the Local Sheet rest frame are plotted as a function of three-dimensional distance from the nearer of the Milky Way or M31.  It is seen that galaxies within 280 kpc of either of the dominant galaxies are usually early type and disperse in velocity.  Galaxies at larger radii are usually late type and manifest a pattern in velocity characteristic of infall.  The pattern suggests a zero velocity (first turnaround) surface at $r_{1t} \sim 940$ kpc.  There are a couple of systems at intermediate radii that depart from the infall
pattern.  NGC 6822 has probably taken a wide swing around the Milky Way and Pegasus has probably passed once by M31.

The discovery that many satellites of M31 are confined to a thin plane with the kinematic signature of rotation \citep{2013Natur.493...62I, 2013ApJ...766..120C} adds an interesting twist to ideas about the formation of the Local Group.   \citet{2013MNRAS.436.2096S} suggest that almost all local satellites lie in four planes.  They construct plausible orbits that suggest the satellites formed in strata as a wall was built around the evacuating Local Void.  Masses are being increasingly constrained as proper motions are measured.

\noindent{\it The Centaurus A complex} is dominated by the peculiar elliptical galaxy NGC~5128 = Cen~A and the beautiful spiral galaxy NGC~5236 = M83.  There are now extensive HST observations of individual galaxies throughout the region, with the result that good distances are known for most systems from Tip of the Red Giant Branch (TRGB) luminosities.  

\begin{figure}[htbp]
\begin{center}
\includegraphics[scale=0.4]{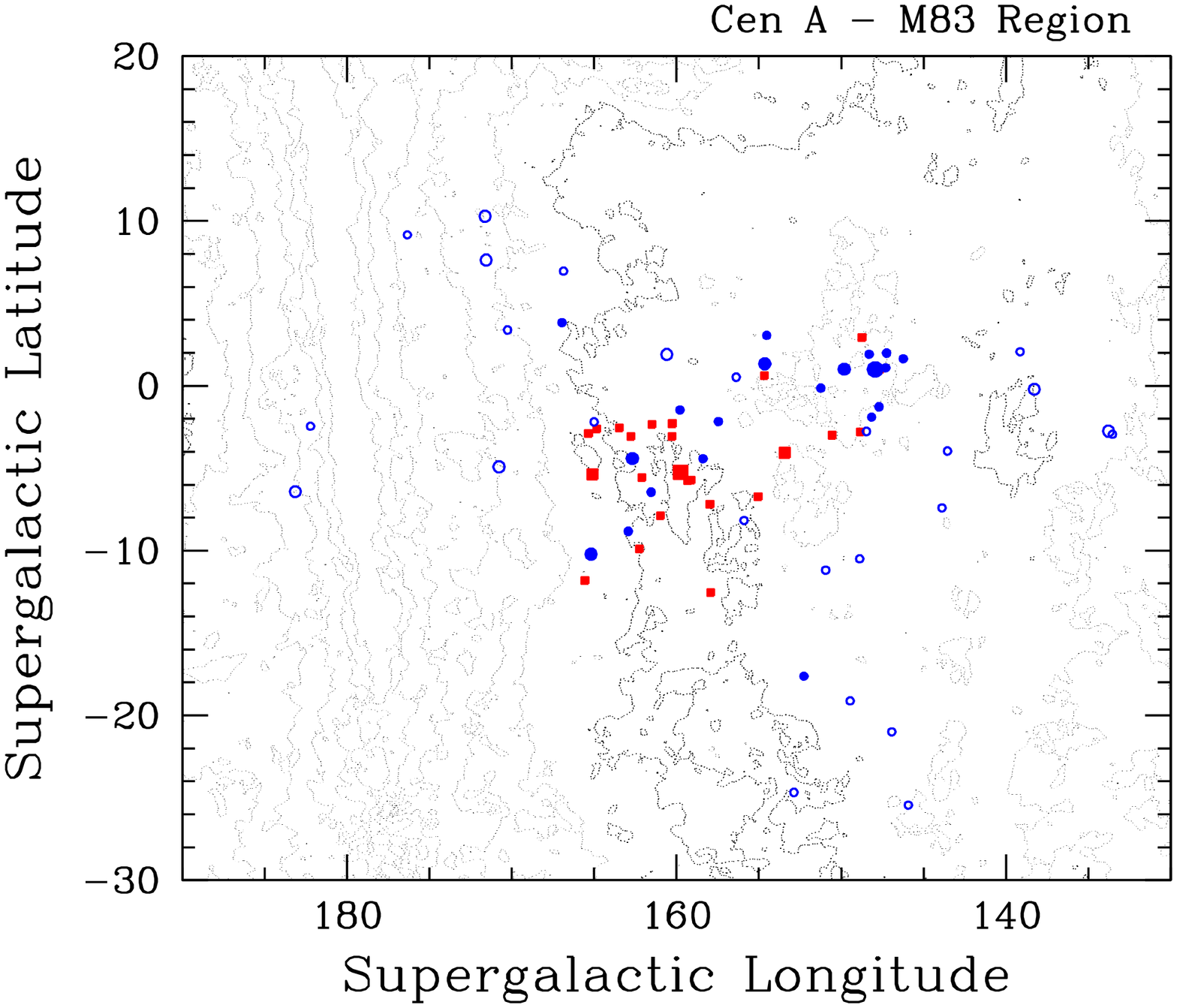}
\caption{Projected distribution of galaxies in the vicinity of Cen~A and M83 in super galactic coordinates.  Early types identified by red squares and late types identified by blue circles.  Filled symbols are within 1 Mpc of either Cen~A or M83 in 3D.   Crosses: uncertain distances.  Obscuration contours are shown at factor 2 increments, with the dark contour at the level $A_B = 0.5$. The Galactic equator lies at roughly SGL=180.} 
\label{cena_complex}
\end{center}
\end{figure}

The structure of the complex has been described by \citet{2002A&A...385...21K, 2007AJ....133..504K}.  Figure~\ref{cena_complex} shows the distribution of galaxies on the sky.  It is seen that there are many early types (Sa and earlier) around Cen~A (SGL=160, SGB=$-5$) and late types (later than Sa) around M83 (SGL=148, SGB=+1).  The Cen~A and M83 families blend in a projection from our vantage point and in their velocities.  We are reminded from the contours of extinction that the Cen~A complex lies near the zone of avoidance and there may be loss of information at SGL$>165$.

\begin{figure}[htbp]
\begin{center}
\includegraphics[scale=0.4]{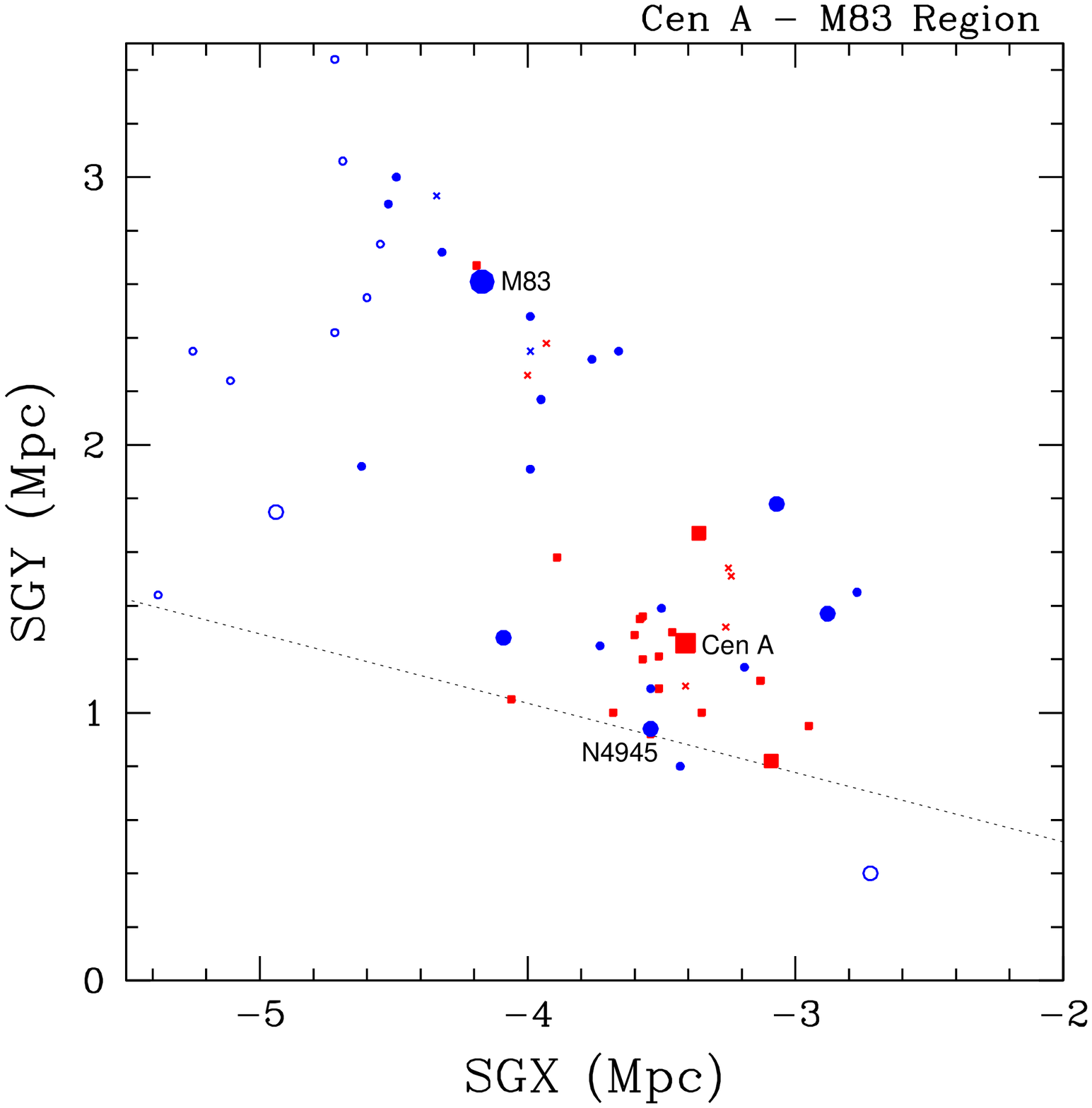}
\includegraphics[scale=0.4]{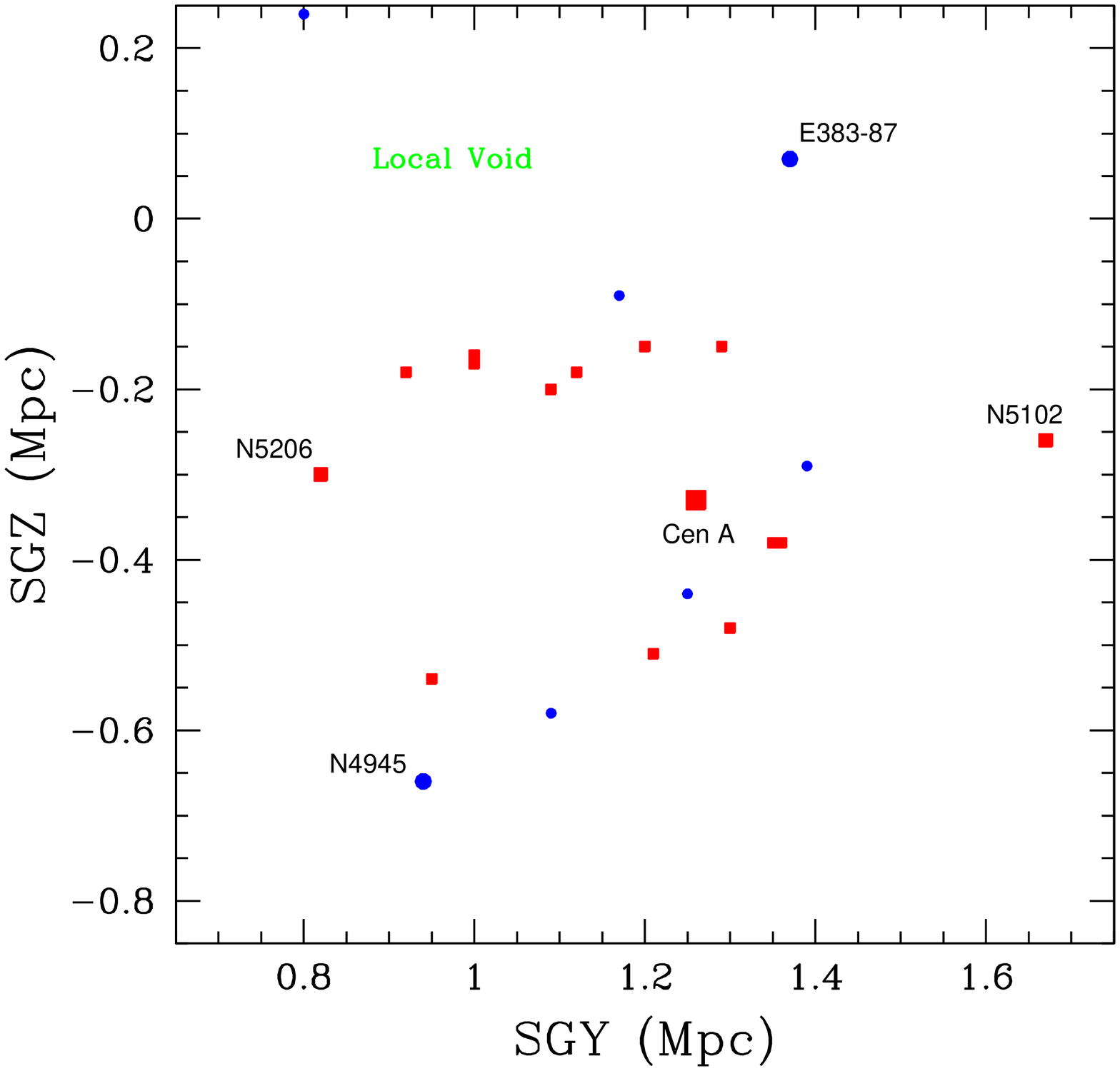}
\caption{{\it Top.} Distribution of galaxies in SGX vs SGY using distance information.  The sample separates into swarms around Cen~A and M83 with a separation between centers of about 1.5 Mpc. {\it Bottom.} Distribution in SGY vs SGZ (edge-on to local structure) of all galaxies in the frame with measured distances within 3.8 Mpc in SGX.
Symbol codes in the two panels have the same meaning as in Fig.~\ref{cena_complex}.  The Local Void extends above the Cen~A complex at positive SGZ.} 
\label{cena_xyz}
\end{center}
\end{figure}

The transition from sky-projection to three-dimensions using distance information in Figure~\ref{cena_xyz} reveals that the complex has two distinct components separated by $\sim 1.5$~Mpc.  The top panel shows the split.  Cen~A is to the foreground at 3.7 Mpc and is surrounded by predominantly early types.  M83 is further away, at 4.7 Mpc, and its companions (at a mean distance of 5.1 Mpc) are mostly late types.

\begin{figure}[!]
\begin{center}
\includegraphics[scale=0.4]{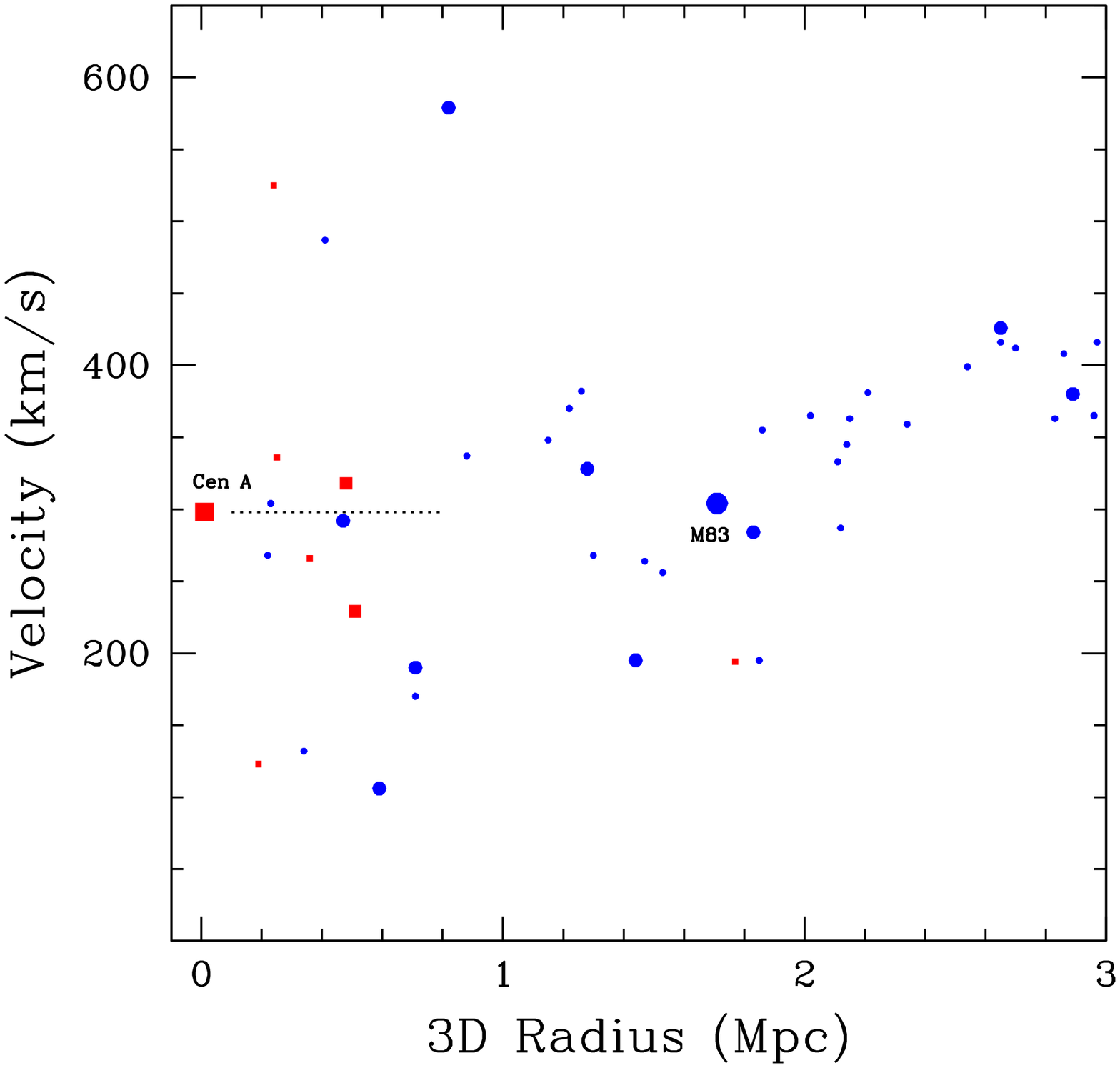}
\caption{Velocities in the Local Sheet rest frame with 3D distance from Cen~A. Red, early; blue, late and the major galaxies are given larger symbols.  The majority of gas-deficient dwarfs do not have known velocities so cannot be included in this figure.  The dashed line at the velocity of Cen~A extends to 800 kpc, the domain of high velocity dispersion around Cen~A.}
\label{cena_r-v}
\end{center}
\end{figure}

The velocities of galaxies in the vicinity are shown in Figure~\ref{cena_r-v} as a function of distance from Cen~A.  Here distances are in 3 dimensions which are more informative but also have larger errors than projected distances.  Within 600 kpc of Cen~A there is a dispersion in 16 velocities of $121\pm30$~\kms\ (unfortunately velocities are  not known for most of the dwarf spheroidals) and the virial mass is $8 \times 10^{12}~\Msun$.  The dispersion of velocities for 9 galaxies around M83 is a smaller $75\pm25$~\kms\ and the inferred virial mass is $2 \times 10^{12}~\Msun$.  The systemic velocities of the two big galaxies and of their groups are the same within the uncertainties.  M83 and its companions are near the zero velocity surface of the larger Cen~A halo.  The two currently distinct groups are on the brink of falling together.

%The predominance and concentration of early types around Cen~A is shown in Figure~\ref{cena_ar-n}.  The family of early types lie within a projected radius of $9^{\circ} = 580$~kpc.  

Given the attention being received by the apparent adherence of satellites around M31 and the Milky Way to thin planes \citep{2013Natur.493...62I, 2013MNRAS.436.2096S} it would be negligent to ignore the apparent occurrence of the phenomenon in the vicinity of Cen~A seen in the lower panel of Figure~\ref{cena_xyz}.  TRGB distances have r.m.s. 5\% uncertainties and every observed galaxy in the vicinity of Cen~A is included in the figure (most of errors in distance project into the SGX axis, normal to the figure; almost none projects into SGZ).  All the galaxies with the possible exception of the most distant cases (NGC 5102 and ESO 383-87) appear to lie in two planes.  Each plane has a long axis of about 600 kpc and a thickness of 100 kpc.  The thickness is reduced slightly from a vantage point about $5^{\circ}$ above the SGX axis but already the pattern is striking in this view along the SGX cardinal axis.  In the case of the local planes, Shaya \& Tully drew attention to the probable role of evacuation from the Local Void in the development of layering of structure.  The proximity of the Local Void is identified in the SGY$-$SGZ plot of Figure~\ref{cena_xyz}. The orientations of the two posited Cen~A satellite planes suggest that they, like their local cousins,  developed out of the evacuation of the void.  

%\begin{figure}[htbp]

%\begin{figure}[htbp]
%\begin{figure}[!]
%\begin{center}
%\includegraphics[scale=0.4]{cena_ar-n3.ps}
%\caption{The cumulative number of early (solid red) and late (dotted blue) galaxies as a function of projected separation from Cen~A (includes all galaxies within 1 %Mpc of Cen~A in 3D).  The radii containing half the respective samples are identified.}
%\label{cena_ar-n}
%\end{center}
%\end{figure}

%\begin{figure}[htbp]
%\begin{figure}[!]
%\begin{center}
%\includegraphics[scale=0.4]{radmpc_cena_loglindeg.ps}
%\caption{Surface density as a function of projected radius from Cen~A.  Only includes galaxies within 1 Mpc of Cen~A in 3D.  Small red triangles: early types; large black circles: all types.}
%\label{cena_radgrad}
%\end{center}
%\end{figure}

%\begin{figure}[htbp]
%\begin{figure}[!]
%\begin{center}
%\includegraphics[scale=0.4]{cena_den2D_nobin.ps}
%\includegraphics[scale=0.4]{cena_den3D_nobin.ps}
%\caption{{\it Top:} Unbinned surface density as a function of projected radius from Cen~A.  Only includes galaxies within 1 Mpc of Cen~A in 3D.  Red triangles: early types; black circles: all types. {\it Bottom:} Unbinned volume density as a function of 3D radius.  Open symbols: nearer M83 than Cen A.
%In both panels, small symbols: individual points; large symbols: smoothed 3-point [1/4, 1/2, 1/4] filter.}
%\label{cena_rad_nobin}
%\end{center}
%\end{figure}

\begin{figure}[htbp]
\begin{center}
\includegraphics[scale=0.4]{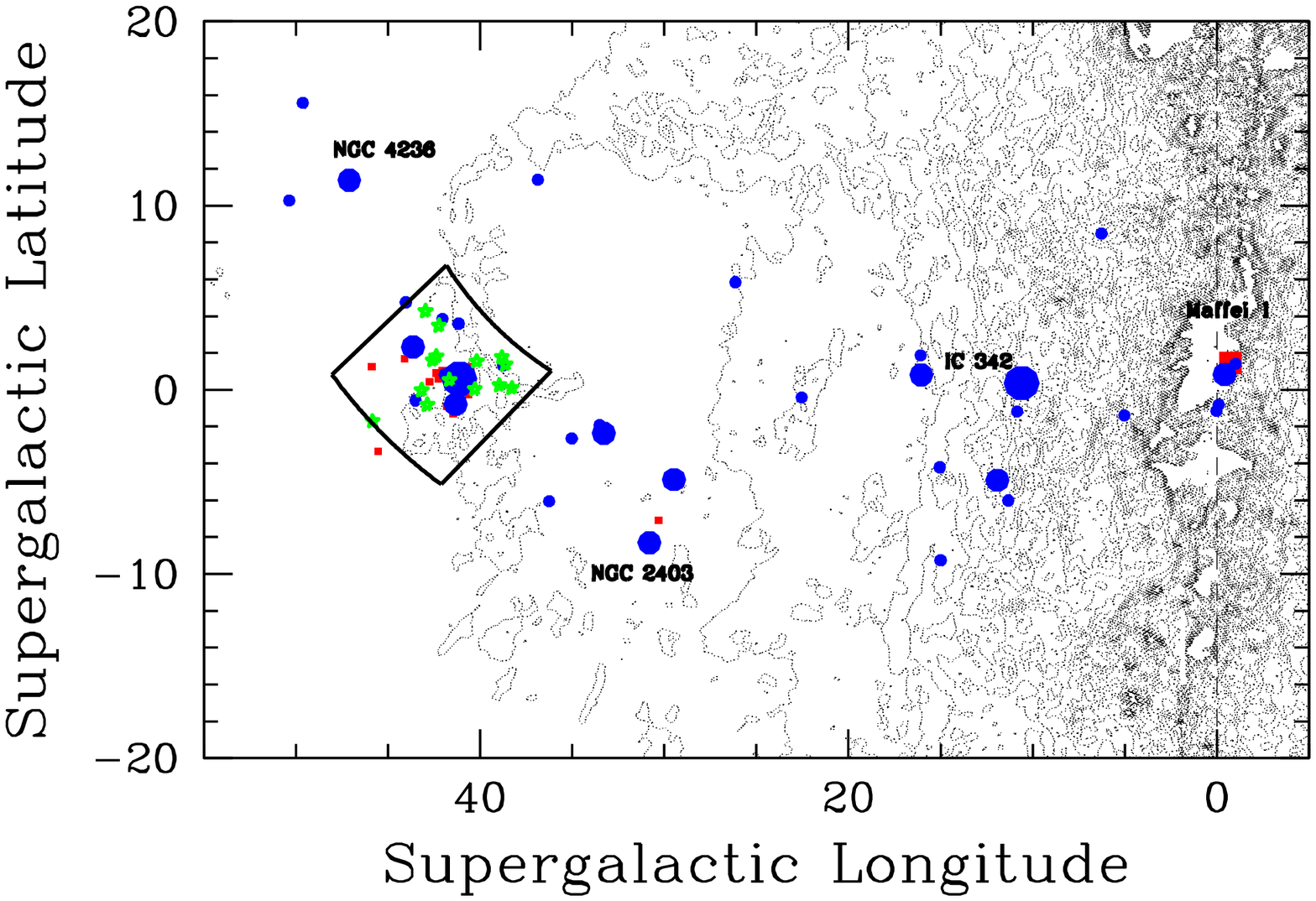}
\includegraphics[scale=0.4]{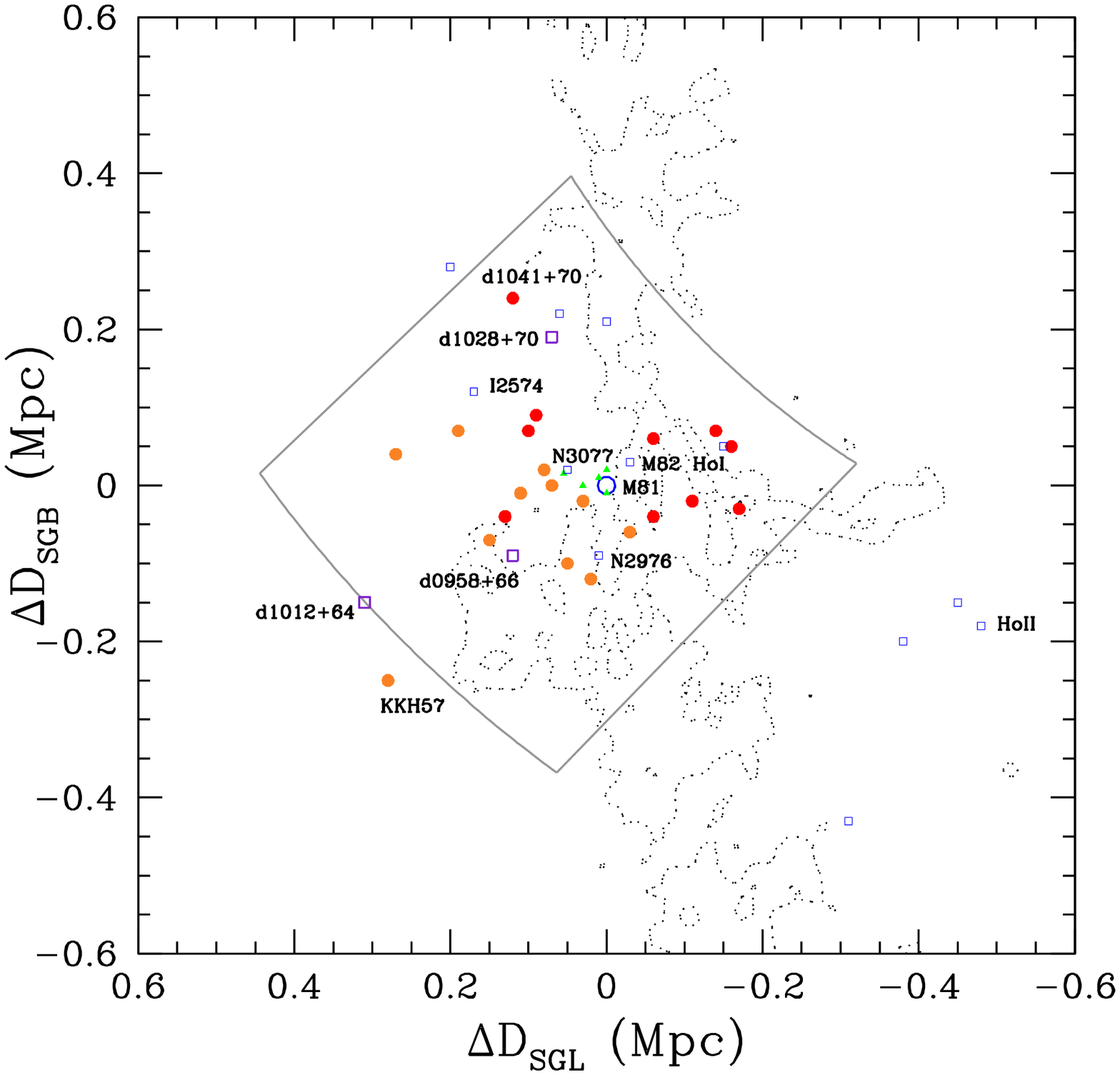}
\caption{
Galaxies in the plane of the sky in the region of M81.  A wide view is shown in the top panel while the bottom panel zooms into the central region of the group.  In the top panel, previously known early types are in red, late types are in blue, and members identified by Chiboucas et al. are in green.  Contours of galactic extinction are given in grey.  The distorted rectangle is an outline of the CFHT imaging survey.  In the bottom panel, early type galaxies that are new or previously known are colored red and orange respectively, likely tidal dwarfs near M81 are represented by green triangles,  while late type systems are located by open squares and three newly identified Blue Compact Dwarfs are labelled. 
}
\label{m81_complex}
\end{center}
\end{figure}

\begin{figure}[htbp]
\begin{center}
\includegraphics[scale=0.4]{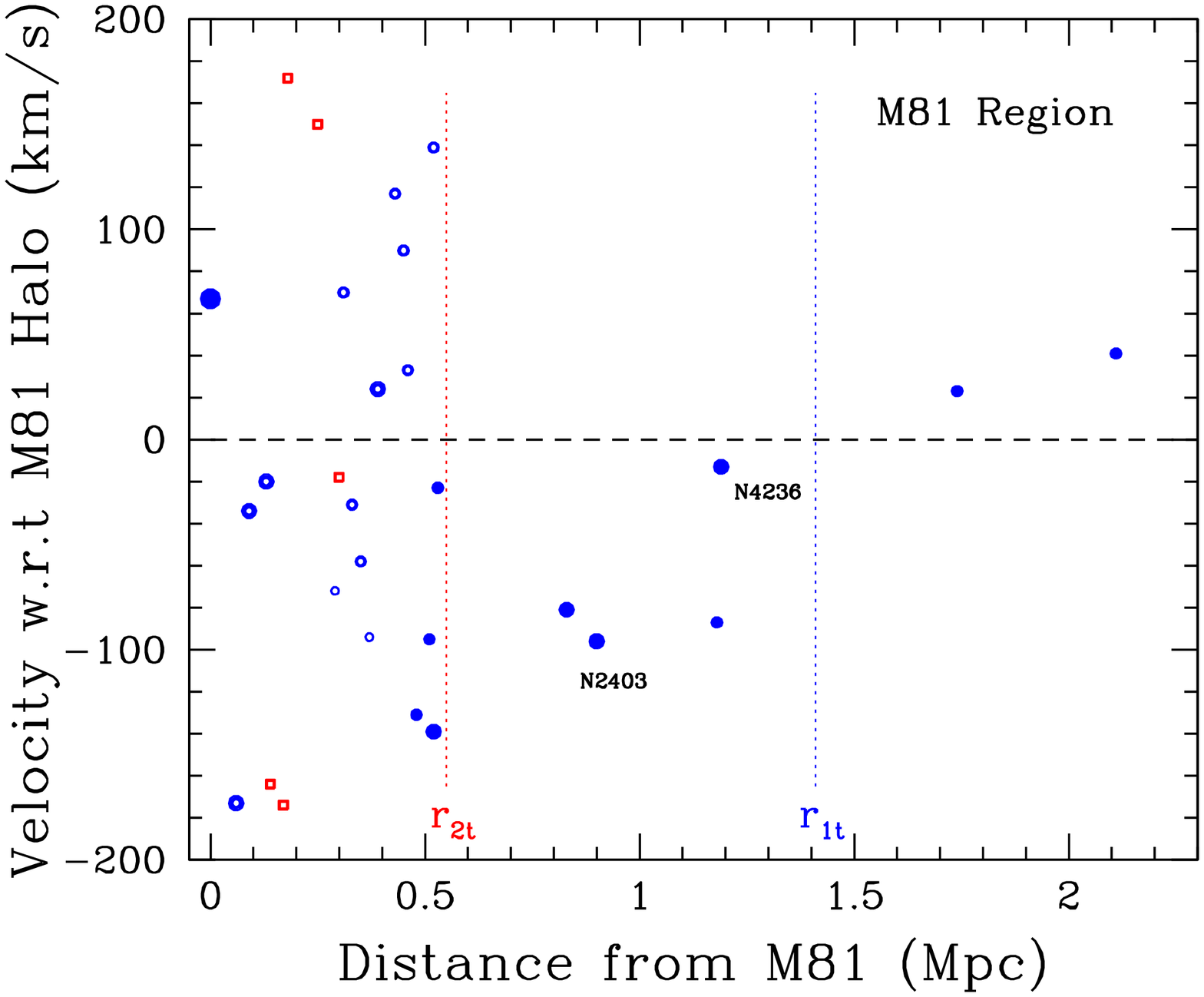}
\caption{
Velocities with respect to the luminosity weighted mean of M81-M82-NGC3077 as a function of 3D distance.  Open symbols: galaxies within a projected radius of 370 kpc of M81.
Galaxies at 3D radii 0.5 -- 1.3 Mpc display the pattern expected of infall.  The vertical dotted lines identify the first and second turnaround radii, $r_{1t}$ and $r_{2t}$.}
\label{m81_rv}
\end{center}
\end{figure}

\noindent {\it The M81 complex} at 3.6 Mpc has traditionally included the region displayed in the top panel of Figure~\ref{m81_complex}.   The nearest important neighbor is IC~342 at a distance of 1.9 Mpc from M81.  IC~342 is part of the Maffei complex, off the right edge of the field shown in Fig.~\ref{m81_complex}, in the direction toward the Galactic plane.  Good distances are now available for the major galaxies in the neighboring IC342/Maffei structure \citep{2014arXiv1404.2987W} but the region is poorly studied because of obscuration so will not be given attention in this discussion.

TRGB measurements based on HST observations \citep{2002A&A...383..125K, 2013AJ....146..126C} provide a detailed three dimensional picture of the M81 region.  It is seen between Figures~\ref{m81_complex} and \ref{m81_rv} that there is a concentration of galaxies around M81 with a substantial dispersion in velocities and there is a domain extending to $\sim 1.3$ Mpc of late type galaxies exhibiting the kinematic characteristics of infall.  Few of the early type dwarfs have known velocities.  The domain of high velocity scatter extends to a 3D radius of 550 kpc, taken to be the radius of second turnaround, $r_{2t}$.  The transition from negative to positive velocities relative to the M81 group velocity centroid occurs at a radius of 1.41 Mpc and is taken to be the radius of first turnaround, $r_{1t}$.   The line-of-sight velocity dispersion from 23 galaxies within the core region is $111\pm23$ km/s and the virial mass is  $3 \times 10^{12}~\Msun$.

%\begin{figure}[htbp]
%\begin{center}
%\includegraphics[scale=0.4]{radgrad_m81_extend_loglin.ps}
%\caption{
%Projected distribution of galaxies with respect to M81; bins of 6 galaxies.}
%\label{m81_radgrad}
%\end{center}
%\end{figure}

%\begin{figure}[htbp]
%\begin{center}
%\includegraphics[scale=0.4]{m81_den2D_nobin.ps}
%\includegraphics[scale=0.4]{m81_den3D_nobin.ps}
%\caption{{\it Top:} Unbinned surface density as a function of projected radius from M81.  Red triangles: early types; black circles: all types. {\it Bottom:} Unbinned volume density as a function of 3D radius.  
%In both panels, small symbols: individual points; large symbols: smoothed 3-point [1/4, 1/2, 1/4] filter.}
%\label{m81_rad_nobin}
%\end{center}
%\end{figure}

\noindent {\it The Sculptor complex} has NGC 253 at its core, 3.6 Mpc from us.  The three dimensional distribution of galaxies in the region are shown in the two projections of Figure~\ref{scl_complex}.  Prior to the availability of good distances, NGC~55 and NGC~300 had been considered part of an NGC~253 group. There is a more recent view that the complex is an extended filament \citep{1998AJ....116.2873J, 2003A&A...404...93K}.

Rather, the excellent distances provided by the TRGB method show conclusively that NGC 55 and NGC 300 are members of an entity that has been called an association of dwarfs \citep{2006AJ....132..729T}.  That association is quite distinct from the NGC~253 Group.  The entity feels a force of attraction from the combination of M31 and our Galaxy that is roughly three times the force from the historical Sculptor group.  This association of dwarfs will be given attention in the next section.

\begin{figure}[!]
\begin{center}
\includegraphics[scale=0.39]{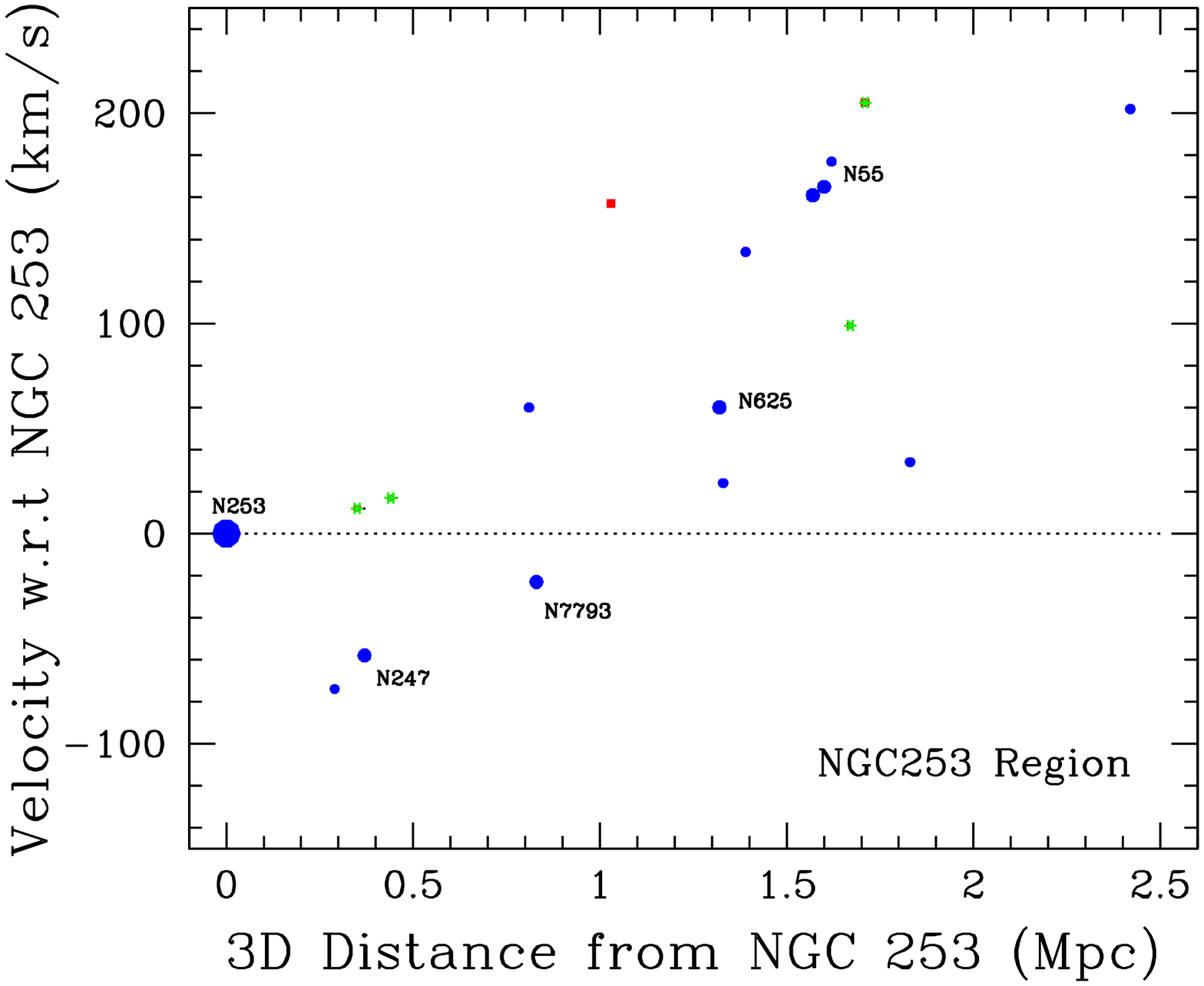}
\caption{
Velocities with respect to the dominant galaxy NGC 253.  There is a hint of an infall region within 800 kpc.}
\label{n253_rv}
\end{center}
\end{figure}

There is evidence for a collapsed region around NGC 253 but the numbers are very small.  In Figure~\ref{n253_rv} it is seen that two or three gas-rich galaxies, including the intermediate-size NGC 247, are within the infall regime.  The velocities of the infalling galaxies provide an envelope for the velocity dispersion so it can be inferred that the velocity dispersion of satellites is at least $\sim 70$ \kms.  The two satellites at slightly positive relative velocities in Fig.~\ref{n253_rv} that lie at 350 and 450 kpc from NGC 253 are transition types, with the morphological appearance of dE but detectable amounts of HI gas \citep{2005AJ....130.2058B}.  These distances  of $400 \pm 50$ kpc from NGC 253 are rather large to be within the second turnaround but the distances are not in doubt because the sky projection components are dominant.   The line-of-sight velocity dispersion of the four nearest galaxies to NGC 253 is 47~\kms, obviously an uncertain number.

\subsection{Associations of Dwarf Galaxies}

Within the radius of 4 Mpc of our position that includes the halos containing Cen~A, M81, and NGC~253 (and the obscured Maffei-IC342 complex) there are four or five entities that appear bound but include only small galaxies \citep{2006AJ....132..729T}.  The characteristic radii of these entities of $\sim 300$~kpc is too large to suppose that they are relaxed.  In the main, it is suspected that these associations are still forming, with motions dominated by infall. 

\begin{figure}[htbp]
\begin{center}
\includegraphics[scale=0.85]{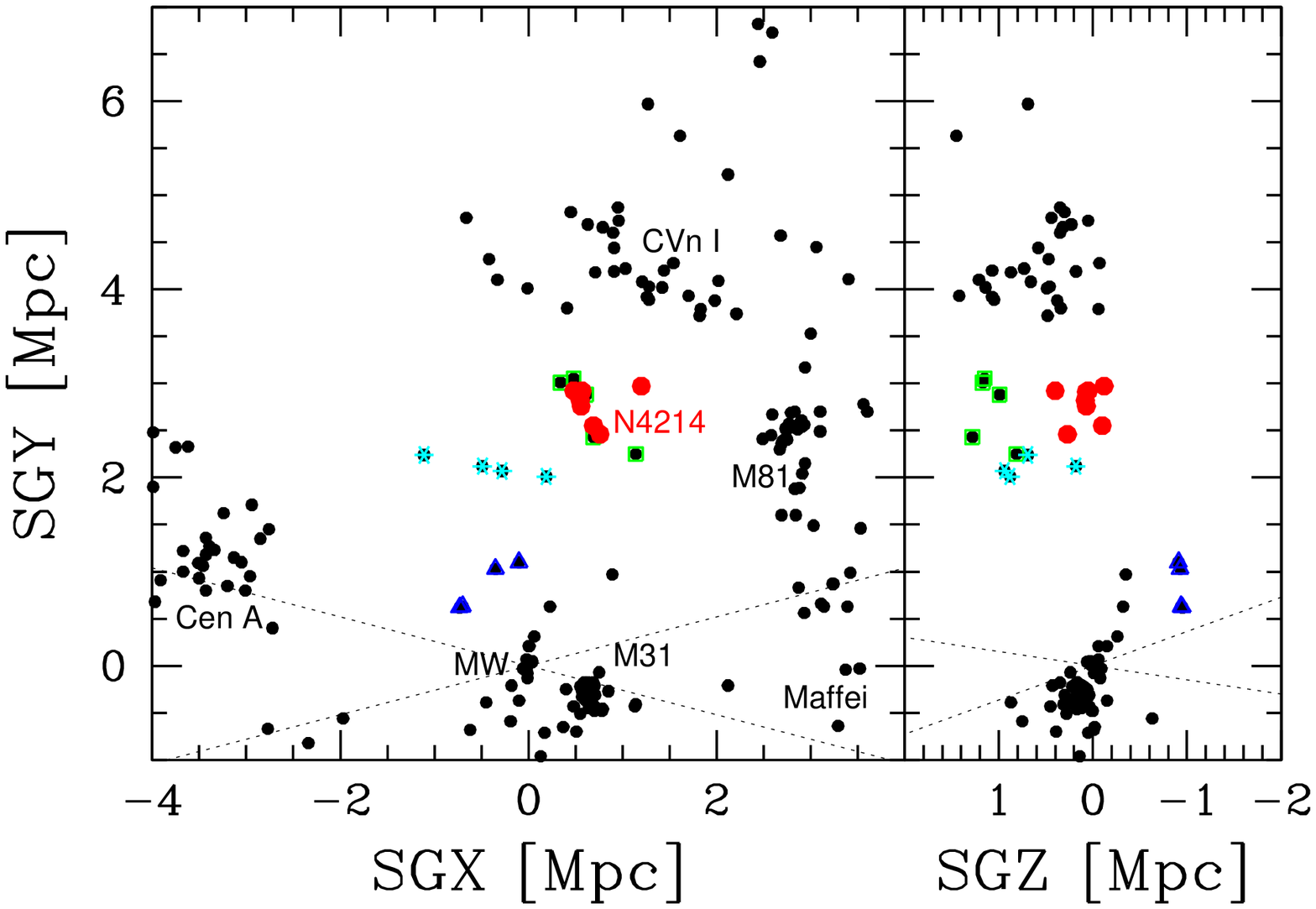}
\caption{
Two supergalactic projections of galaxies in the NGC 4214 association (larger red filled circles).  The adjacent complexes include 3 other dwarf associations: NGC 3109 (blue triangles), 14+8 (green enclosing boxes), and "dregs" (cyan stars).}
\label{n4214}
\end{center}
\end{figure}

In two of these nearby cases, the process of collapse seems sufficiently advanced and the numbers of participants are sufficient to warrant a brief analysis.  One case involves the $M_B = -17.1$ Magellanic irregular galaxy NGC~4214 as its brightest member and includes 8 known galaxies within 3D 600 kpc (adding NGC~4190 and UGC~7949 to the galaxies identified with the 14+7 group in \citet{2006AJ....132..729T}).  This association overlaps in sky projection and velocity with the Canes Venatici I group at 4.7~Mpc but good TRGB distances confirm the association of dwarfs lies at 2.9 Mpc.  The 3D relationship of the 8 small galaxies is shown in the two panels of Figure~\ref{n4214}.  The association has a core, with four galaxies including NGC~4214 within an enclosed domain of $\sim 100$~kpc radius.  The velocity dispersion for these four galaxies is $\sigma_p = 49$~\kms\ and the virial mass estimate is $3 \times 10^{11}~M_{\odot}$.  Including all 8 galaxies in the analysis, the bi-weight velocity dispersion is $51\pm18$~\kms, the 3D gravitational radius doubles to 435 kpc, and the virial mass increases to $7 \times 10^{11} M_{\odot}$.   The luminosity at $B$ band is only $1.4 \times 10^9 L_{\odot}$ so the implied mass-to-light ratio is high: $M/L_B \sim 230$ to $470 M_{\odot}/L_{\odot}$.

\onecolumn
\begin{figure}[htbp]
\begin{center}
\includegraphics[scale=.9]{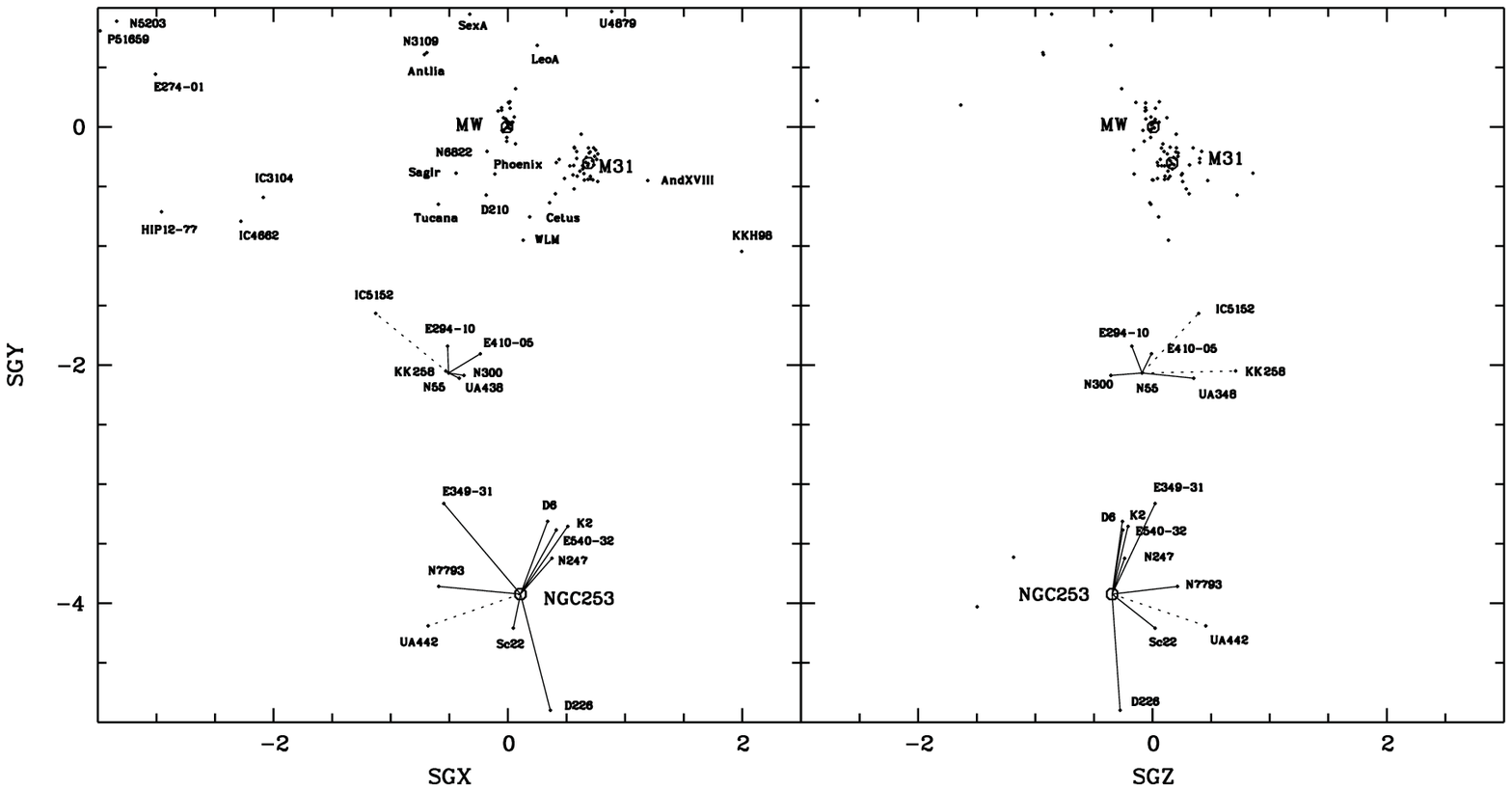}
\caption{Two supergalactic projections of galaxies in the Sculptor direction.  The principal concentration is around the dominant galaxy NGC 253.  There is a secondary concentration associated with NGC 55.  The proximity to the Milky Way and M31 galaxies is seen.}
\label{scl_complex}
\end{center}
\end{figure}
%\twocolumn

\twocolumn

\onecolumn 
\twocolumn

Figure~\ref{n4214} shows the relationship between the NGC4214 association and other nearby galaxies.  Color and special symbols draw attention to three other smaller associations of dwarfs in the vicinity discussed by \citet{2006AJ....132..729T}.  The NGC~3109 association of 4 galaxies (14+12 group) is nearest the Milky Way.  Then quite near the NGC~4214 group are four dwarf galaxies identified with the 14+8 group and 4 more galaxies strung out over a sufficient length that they are referred to as `dregs' in \citet{2006AJ....132..729T}.

The other case worth attention involves five closely associated galaxies and two outliers including NGC~55 ($M_B = -17.8$, type Sdm) and NGC~300 ($M_B = -17.7$, type Sd).  The proximity of these galaxies to NGC~253 was shown in Fig.~\ref{scl_complex} but M31 and our Galaxy have the greater gravitational influence.   The velocity dispersion for the 7 galaxies is $37\pm14$ \kms\ and virial mass estimates, whether the inner 5 or the full 7, range over  $3-4 \times 10^{11}~M_{\odot}$.

%\begin{figure}[htbp]
%\begin{center}
%\includegraphics[scale=0.39]{n4214gp_rv.ps}
%\caption{
%Velocities with respect to NGC 4214.  Horizontal lines reflect $1 \sigma$ uncertainties in line-of-sight distances.}
%\label{n4214_rv}
%\end{center}
%\end{figure}

%\begin{figure}[htbp]
%\begin{center}
%\includegraphics[scale=0.39]{n55gp_rv.ps}
%\caption{
%Velocities with respect to NGC 55.}
%\label{n55_rv}
%\end{center}
%\end{figure}

%Table 1 assembles results for 8 groups.  The three cases discussed above and two others derive from the CFHT Megacam program. Data for the other three -- the Local Group, Virgo Cluster, and Coma Cluster -- are extracted from the literature.[plots from Virgo or Coma analysis???]

\section{Scaling Relations}

As a prologue to the ensuing discussion, the reader is reminded that dimensions and masses have been calculated from direct measurements of distances without recourse to systemic velocities and an assumption about the Hubble Constant.  Of course, the distance scale is established by a zero point set by nearby calibrators.  The distances are compatible with H$_0 = 75$~\kmsMpc\ \citep{2012ApJ...758L..12S}, so rescaling that would alter the choice of this parameter is monitored by $h_{75} = {\rm H}_0/75$.

\begin{figure}[htbp]
\begin{center}
\includegraphics[scale=0.4]{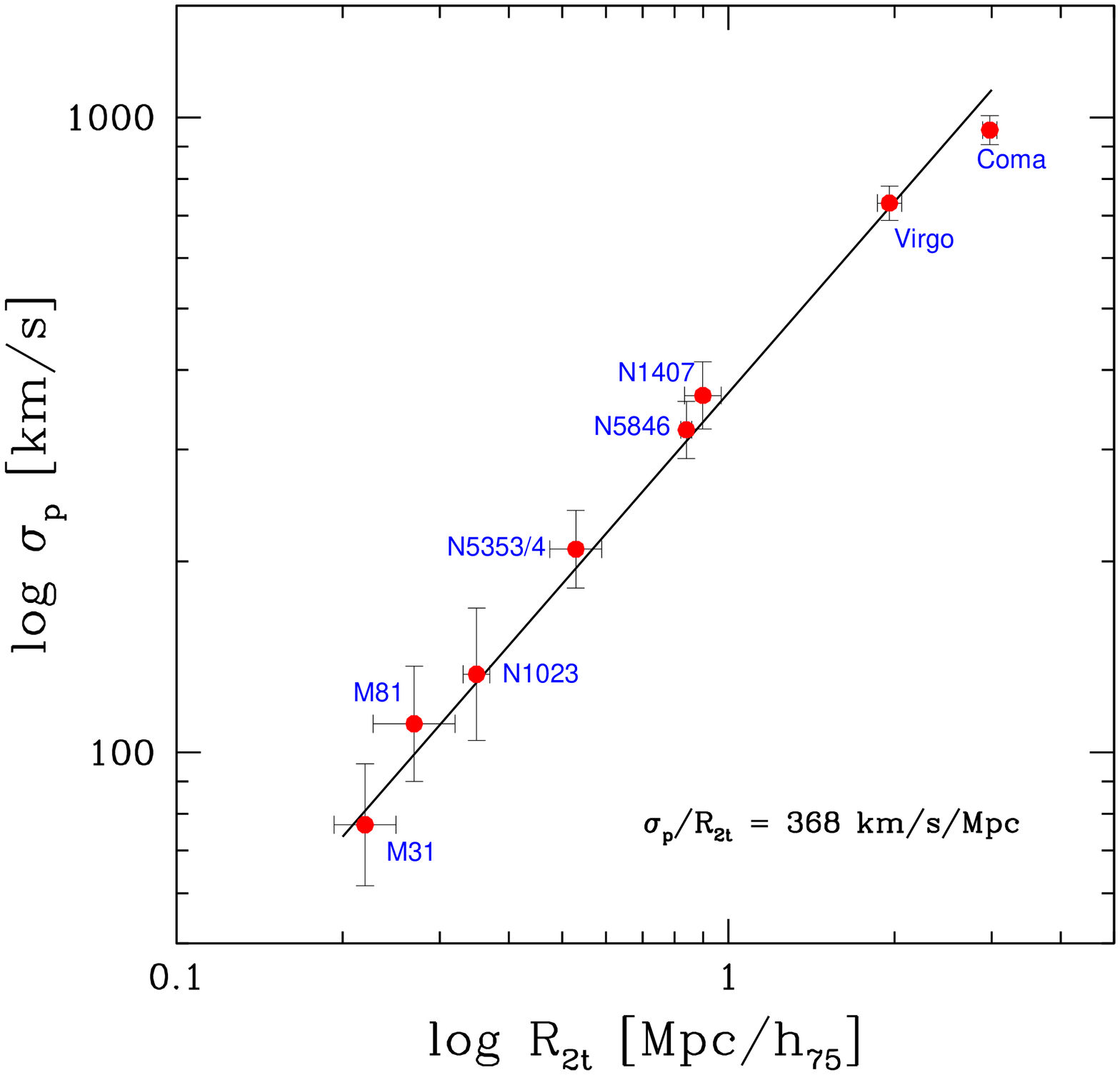}
\caption{Correlation between projected second turnaround radius $R_{2t}$ and the radial velocity dispersion of galaxies within this radius $\sigma_p$.}
\label{rv}
\end{center}
\end{figure}

It was anticipated in the introduction that the two independent variables $R_{2t}$ and $\sigma_p$ would be directly correlated.  The relationship is seen in Figure~\ref{rv}.  The constant of proportionality is found empirically to be $368~h_{75}\pm8$ \kmsMpc.

\begin{figure}[htbp]
\begin{center}
\includegraphics[scale=0.4]{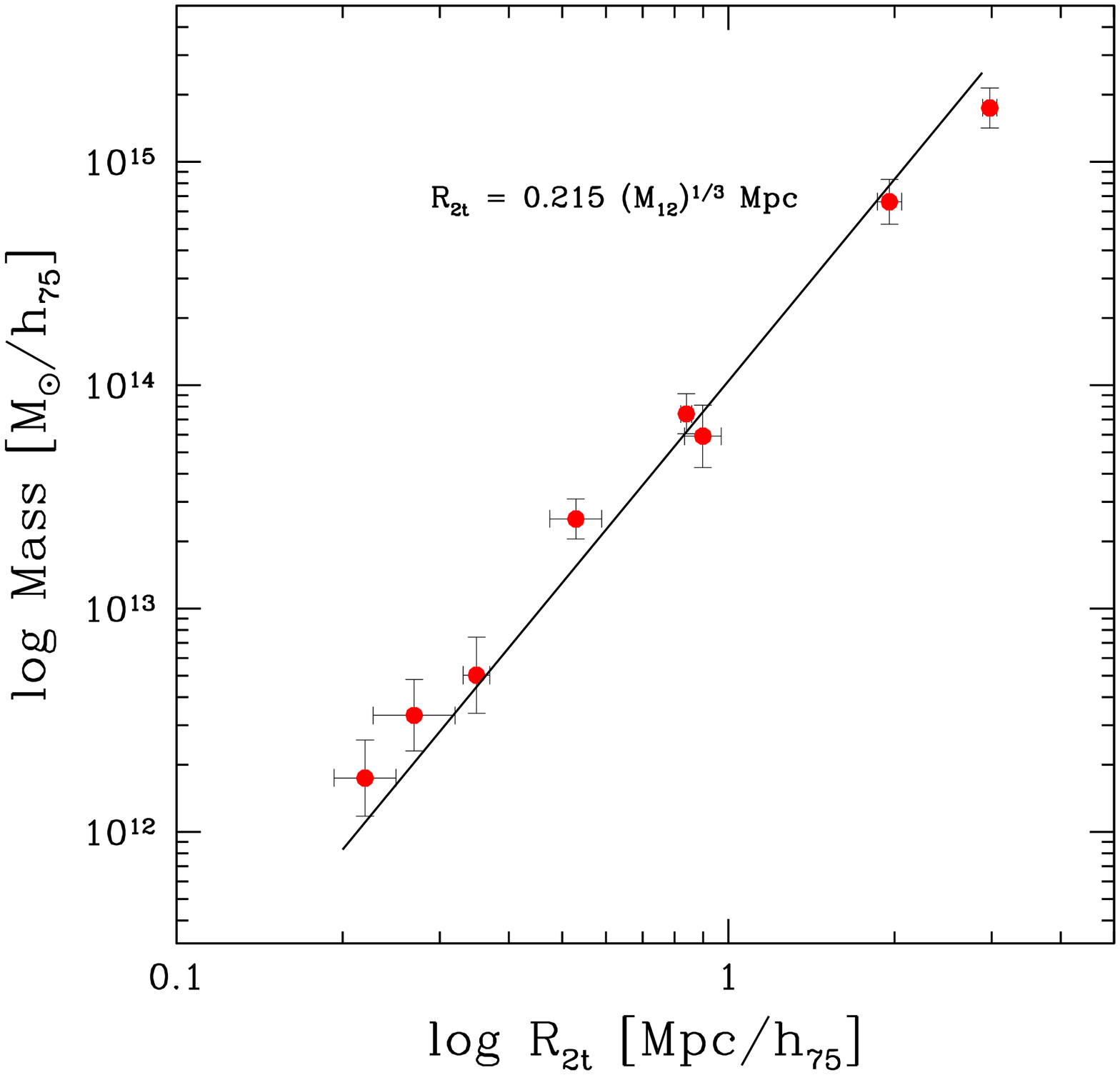}
\caption{Correlation between the projected second turnaround radius $R_{2t}$ and the virial mass calculated for galaxies within this radius $M_{\rm v}$.}
\label{rm}
\end{center}
\end{figure}

\begin{figure}[htbp]
\begin{center}
\includegraphics[scale=0.4]{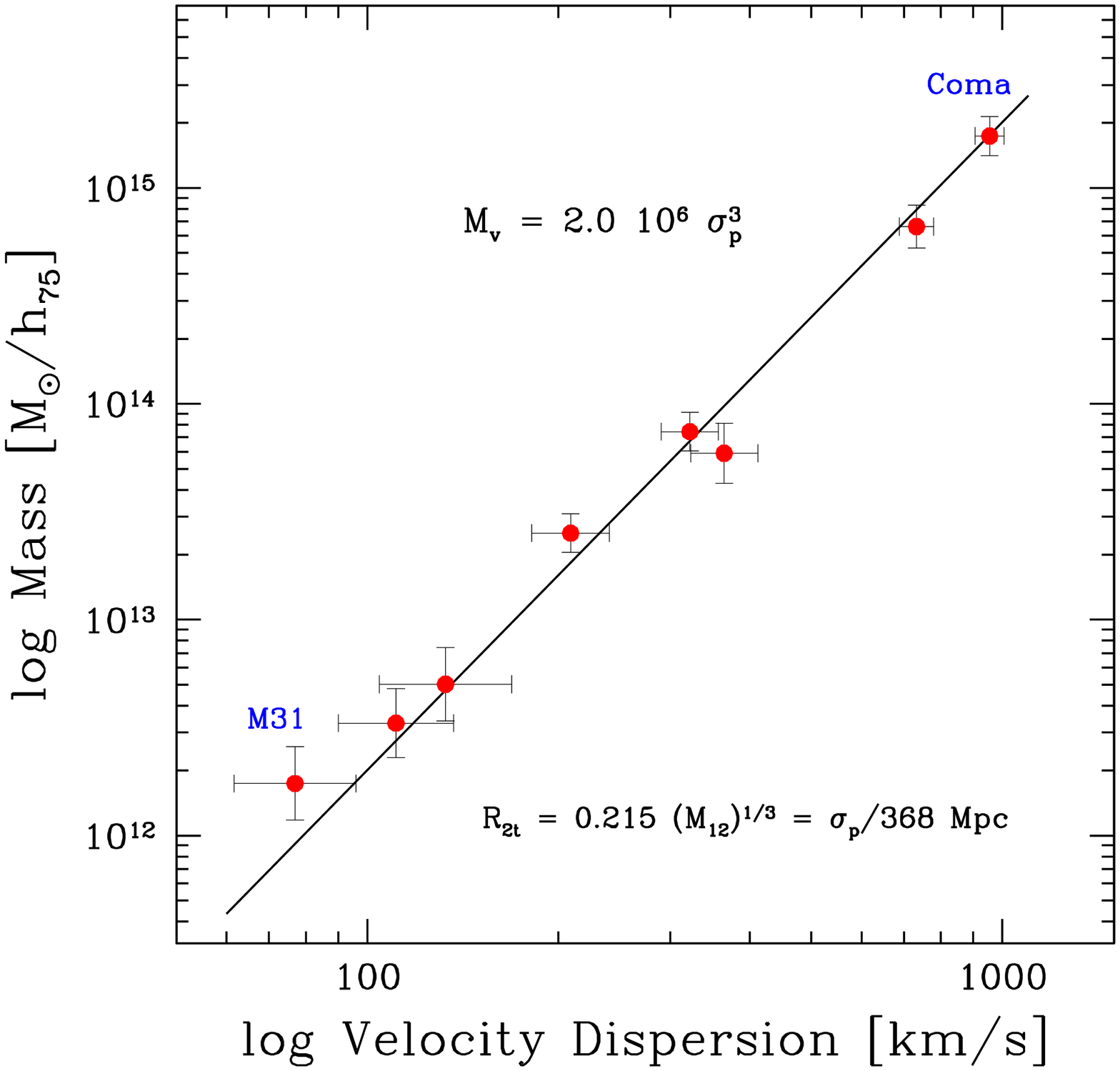}
\caption{Correlation between the virial mass and the velocity dispersion of galaxies within $r_{2t}$.}
\label{mv}
\end{center}
\end{figure}

Two interesting relations are shown in Figures \ref{rm} and \ref{mv}.  Each involves the virial mass calculated according to
\begin{equation}
M_{\rm v} = {{\sigma_{3D}^2 r_g} \over {G}} = {{\alpha\pi \sigma_p^2 R_g}\over{2 G}}~.
\end{equation}
The parameter $\alpha$ that describes the nature of orbits is given the value $\alpha = 2.5$ in accordance with the discussion following Eq.~\ref{eq:Mv}.
The gravitational radius $r_g = (\pi/2) R_g$ was defined in Eq.~\ref{rg}
so is calculated independently from $r_{2t} = \sqrt{3/2} R_{2t}$.  What is being plotted in Fig.~\ref{rm} is $R_{2t} \propto \sigma_p^{2/3} R_g^{1/3}$.

The derived correlation is 
\begin{equation}
R_{2t} = 0.215~M_{12}^{1/3}~h_{75}^{-2/3}~{\rm Mpc} 
\label{eq:r2t}
\end{equation}
if the virial mass is in units of $10^{12}~\Msun$.   The standard deviation of the mean fit for the coefficient is $0.006$.   Statistically accounting for projection, the 3D formulation is $r_{2t} = \sqrt{3/2} R_{2t} = 0.263~M_{12}^{1/3}~h_{75}^{-2/3}$ Mpc.  By comparison, the radius $r_{200}$ that encloses a mean density 200 times the critical density is $r_{200} = 0.198~M_{12}^{1/3}~h_{75}^{-2/3}$, so $r_{2t} = 1.33~r_{200}$.

The tight correlation implies the close correlations between projected dimensions $R_{2t} = 1.22 R_g$ and statistically derived 3D dimensions $r_{2t} = 0.95 r_g$. 
It follows that the correlation seen in Fig.~\ref{mv} would exist.  In this plot the slope is fixed and the zero-point is determined by the relations seen in Figs. \ref{rv} and \ref{rm}.  The mass--dispersion relation for groups obeys the law
\begin{equation}
M_{\rm v} = 2.0 \times 10^6~\sigma_p^3~h_{75}^{-1} \Msun
\end{equation}
where mass $M_{\rm v}$ is in solar units and velocity dispersion $\sigma_p$ is in \kms.

\section{The Radius of First Turnaround}

\begin{figure}[htbp]
\begin{center}
\includegraphics[scale=0.3]{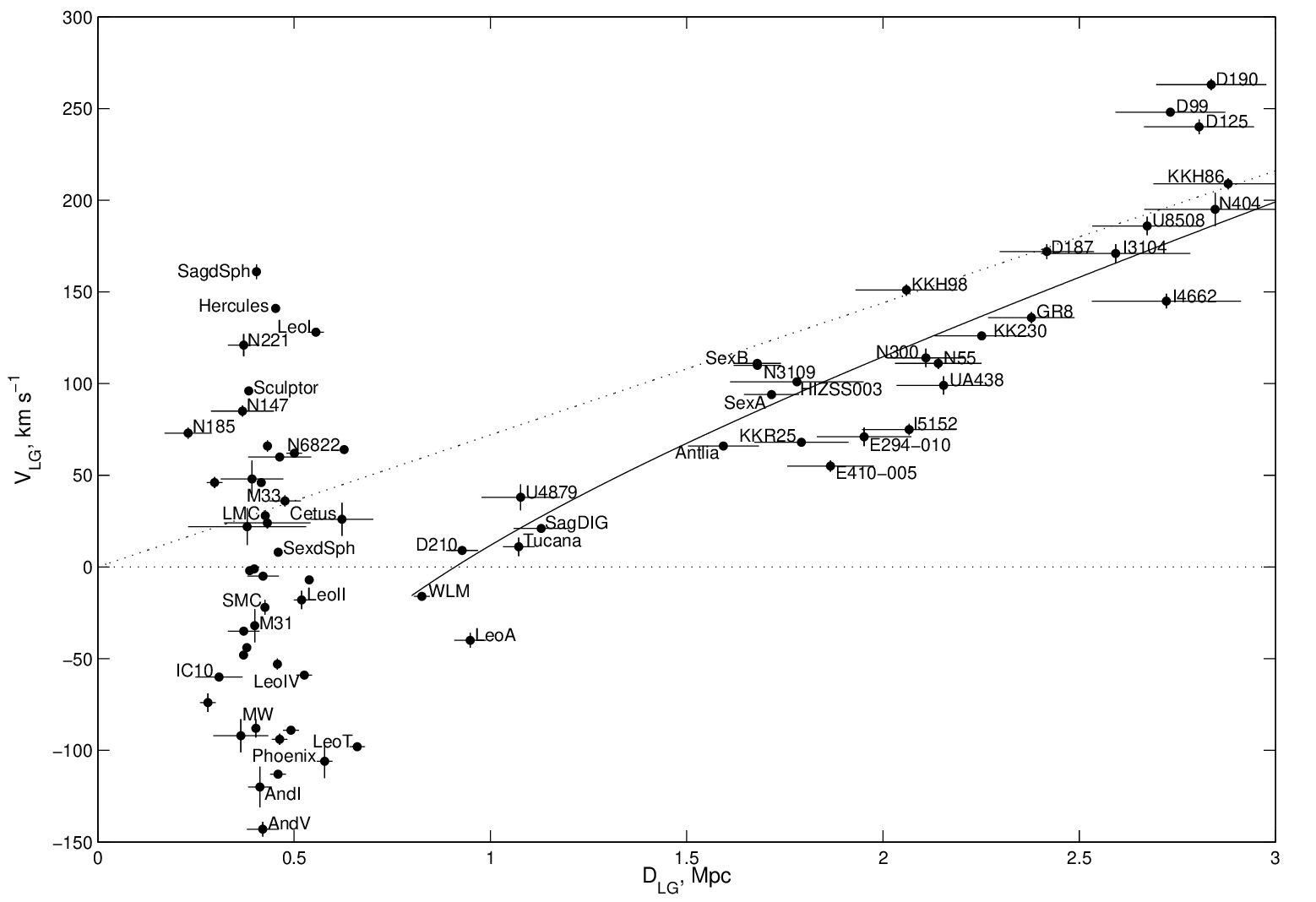}
\caption{Velocities of nearby galaxies in the Local Group rest frame.  The nearest galaxies have a large velocity dispersion.  Galaxies beyond 940~kpc are systematically redshifted.  The solid curve is an empirical fit to galaxys beyond 900~kpc.}
\label{zv}
\end{center}
\end{figure}

A measurement of the radius of first turnaround $r_{1t}$ requires 3D information to distinguish the decoupling of the infall zone from cosmic expansion.  Such information is currently available in only a few cases.  In the case of the traditional Local Group,
an initial discussion by    \citet{1986ApJ...307....1S} was carried forward most recently by  \citet{2009MNRAS.393.1265K}.  Graphical results drawn from that reference are shown in Figure~\ref{zv}.  The value of $r_{1t} = 940$ kpc displayed graphically in Figure~\ref{MW-M31} seems reasonably constrained to an uncertainty of 10\% and encloses a mass of roughly $3 \times 10^{12}~\Msun$.  However spherical symmetry is a poor approximation for the zero velocity surface of the Local Association because of the dumbbell distribution of mass between M31 and the Milky Way.  Moreover, only 6 galaxies are well placed to define the zero velocity surface.

The M81 Group offers a second opportunity to locate a zero velocity surface.  It is seen in Figure~\ref{m81_rv} that $r_{1t} \sim 1.4$ Mpc for this group with $\sim 3 \times 10^{12} \Msun$.    Then a third case is given by the Virgo Cluster where $\sim 8 \times 10^{14} \Msun$ is enclosed within a zero velocity surface $r_{1t} = 7.2\pm0.7$ Mpc \citep{2014ApJ...782....4K}.

This limited information gives us a preliminary calibration of the relationship between $r_{1t}$ and the mass within this radius:
\begin{equation}
r_{1t} = 0.77~M_{12}^{1/3}~h_{75}^{-2/3}~{\rm Mpc}.
\label{eq:r1t}
\end{equation}
with a mean deviation of $\pm0.07$ with three cases.

This empirical determination can be compared with a simplification of the theoretical dependency for spherical collapse (Gary Mamon, private communication):
\begin{equation}
r_{1t}^{theory} = 0.75~M_{12}^{1/3}~h_{75}^{-2/3}~(\Omega_{\Lambda}/0.7)^{2/9}~{\rm Mpc}.
\label{eq:r1t_theory}
\end{equation}

\section{What is a Group?}

It was asked in the introduction how observers should identify groups.  Most usefully, it should approximate the definitions given by modelers.  The parameter $r_{2t}$, the radius of second turnaround, is very close to the parameter $r_{200}$, the radius enclosing an overdensity of 200 times closure density, or $r_g$, the gravitational radius. 
A modeler identifies collapsed halos with dimensions that resemble $r_{2t}$.  This is a dimension commonly associated with rich clusters.  If it is accepted that the terms `group' and `cluster' are interchangeable, then the dimension an observer can use to define collapsed structure is $r_{2t}$. 

By this measure, the Local Group is two groups, one around M31 and one around the Milky Way.  Mass derivations result in similar estimates for each of MW and M31 of $1.5 \times 10^{12}~\Msun$.  It would follow that $r_{2t} \sim 280$ kpc for both groups.  The pair are falling together, hence lie within a common first turnaround surface enclosing $\sim 3 \times 10^{12}~\Msun$.  Figure~\ref{1Mpc} illustrates the projected positions of galaxies in the region and it is seen that satellites of early type are predominant within 280 kpc of M31 and the Milky Way while satellites of late type are predominant at larger radii.  Figure~\ref{MW-M31} shows radial velocities and it is seen that satellites within 280 kpc of one of the major galaxies have large motions, positive and negative, while the outlying galaxies mostly display the characteristics of infall.

The radius of first turnaround, $r_{1t}$ can be a useful construct.  It is proposed that entities defined by this radius be called `associations'.  We live in the Milky Way Group in the Local Association.  Of course, since they are within a common infall envelope, the Milky Way and M31 halos are destined to merge.  The current estimate is that first periastron is 3.9 Gyr in the future  \citep{2012ApJ...753....9V}.
It turns out that most other nearby structures that have been called groups break up into multiple parts.  The M81 Association contains the M81 Group, the NGC 2403 Group, and a common infall region.  The Centaurus Association contains the Cen A Group and the M83 Group as major components.  There is the Maffei--IC 342 Association.   

The `Main Disturber' tidal index in the {\it Updated Nearby Galaxies Catalog} \citep{2013AJ....145..101K} provides a description that approximates the definition of association given here.   \citet{2011MNRAS.412.2498M} have compiled a catalog of groups that are reasonably consistent in scale with a second turnaround definition.
The confusion in the definition of the word `group' is implicitly acknowledged in the study by \citet{2005AJ....129..178K} of the nearby extragalactic structure.  In that work, traditional `groups' are referred to as complexes or filaments or clouds and the collapsed regions are called groups. 

\section{Dark Matter, Dark Energy, and Local Dynamics}

The relationships between $r_{2t}$ or $r_{1t}$ and the masses internal to those radii are virtually independent of the dark energy content of the universe.  However, because of the relative differences with respect to the age of the universe between the onset of collapse that is associated with $r_{2t}$ and the current age associated with $r_{1t}$, the ratio $r_{1t}/r_{2t}$ has a dependence on the dark energy content.

The equations \ref{eq:r2t} and \ref{eq:r1t} can be combined, after adjustment for projection, to give
\begin{equation}
r_{1t}/r_{2t} = 3.14\pm0.28 .
\end{equation}
This ratio can be compared with a theoretical prediction, never published but kindly provided by Gary Mamon.  The theoretical analysis involves the virial radius in place of the second turnaround radius. 
The ratio of the two turnaround radii as a function of matter density are calculated by Mamon using alternative approximations of the virial density by \citet{1996MNRAS.280..638K} and   \citet{1998ApJ...495...80B}.  \citet{2005MNRAS.361L...1W} provide a link between virial and second turnaround radii.  They define the virial radius to enclose a region with 101.9 times the critical density.  They then determine the following relation from simulated halos in the mass range $10^{14}-10^{15}~\Msun$:
\begin{equation}
r_v = 0.245~M_{12}^{1/3}~h_{75}^{-2/3} {\rm Mpc}.
\end{equation}
This formulation is very close to the 3D adjusted version of Eq.~\ref{eq:r2t} for the second turnaround radius.  Hence $r_{2t} \simeq 1.07 r_v$.

\begin{figure}[htbp]
\begin{center}
\includegraphics[scale=0.39]{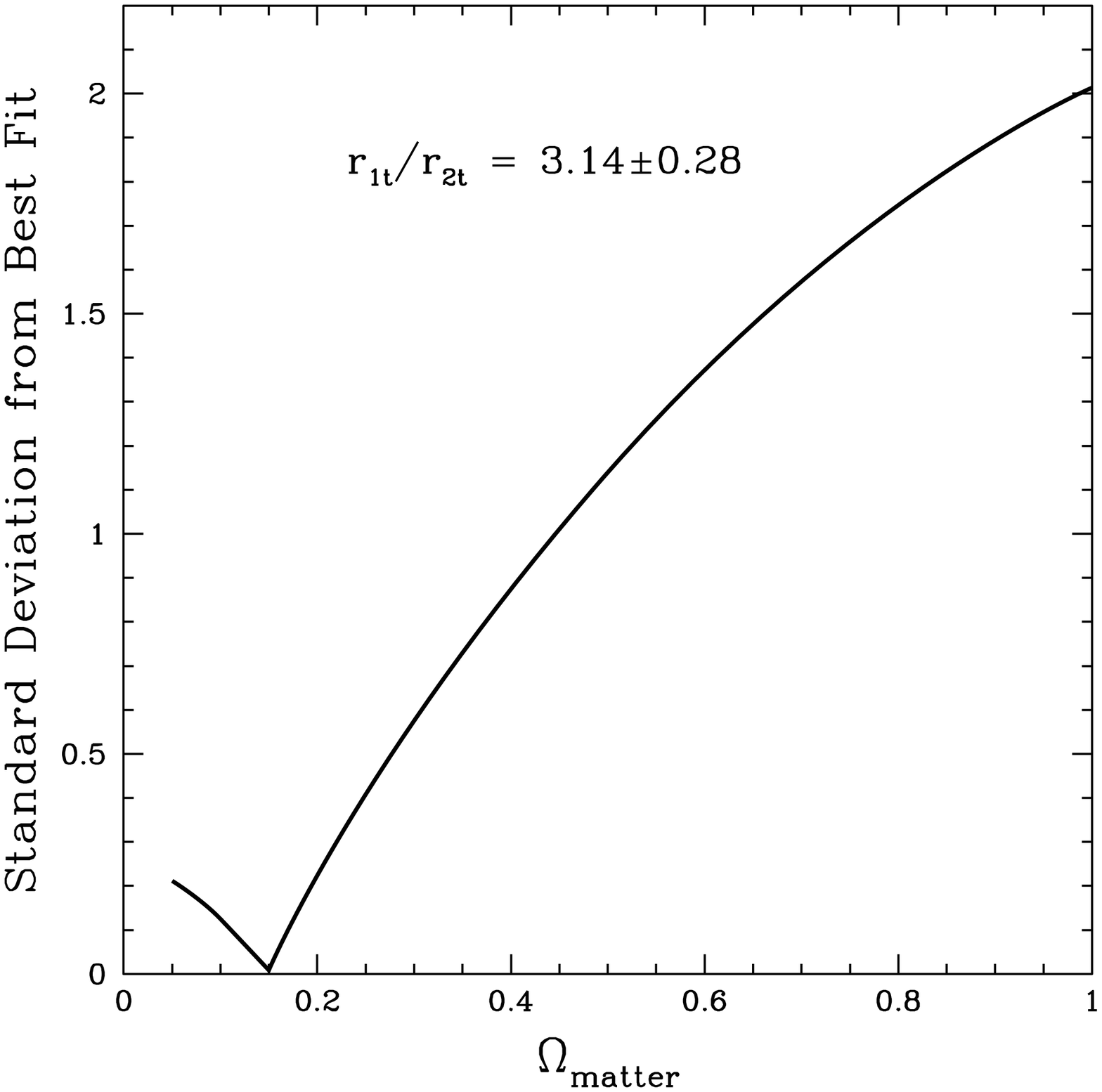}
\caption{Measure of agreement between observed ratio of first and second turnaround radii and expectation value as a function of matter density in a universe with flat topology.  The best fit value of $r_{1t}/r_{2t}=3.14$ implies $\Omega_{matter} \sim 0.15$ while at a deviation of $+1 \sigma$ the correspondence is $\Omega_{matter} \sim 0.44$.}
\label{ta}
\end{center}
\end{figure}

Assuming the sum of matter and vacuum energy densities add to the critical density, the ratio $r_{1t}/r_{2t}$ ranges from 3.1 if matter density is extremely low to 3.7 in the Einstein-de Sitter case.  The observed ratio of 3.14 corresponds to the case $\Omega_{matter} = 0.15$.
Figure~\ref{ta} illustrates the viability of alternative density parameters.   
The 68\% probability domain is constrained to $\Omega_{matter} < 0.44$. However, $\Omega_{matter} = 1$ is only rejected with a significance of $\sim 2 \sigma$.      
We have here an interesting potential test of the local influence of dark energy but the current uncertainties are very large. 

Another quantity of potential interest is the surface of zero gravity, the surface around an overdense region that is at the limit of what is destined to ever collapse onto the overdensity.  This surface is well specified as a function of cosmological model in the case of spherical collapse.  With $\Omega_{\Lambda} = 0.7$ in a flat universe then $r_{ZG} \sim 1.4 r_{1t}$   \citep{2008A&A...488..845P}.  However it was mentioned already with the discussion of $r_{1t}$ that spherical collapse is generally a bad approximation because of the neghboring structures.  It must be a very bad approximation at $r_{ZG}$.

\section{Evolved Groups, Spiral Groups, and Dwarf Associations}

\begin{figure}[htbp]
\begin{center}
\includegraphics[scale=0.39]{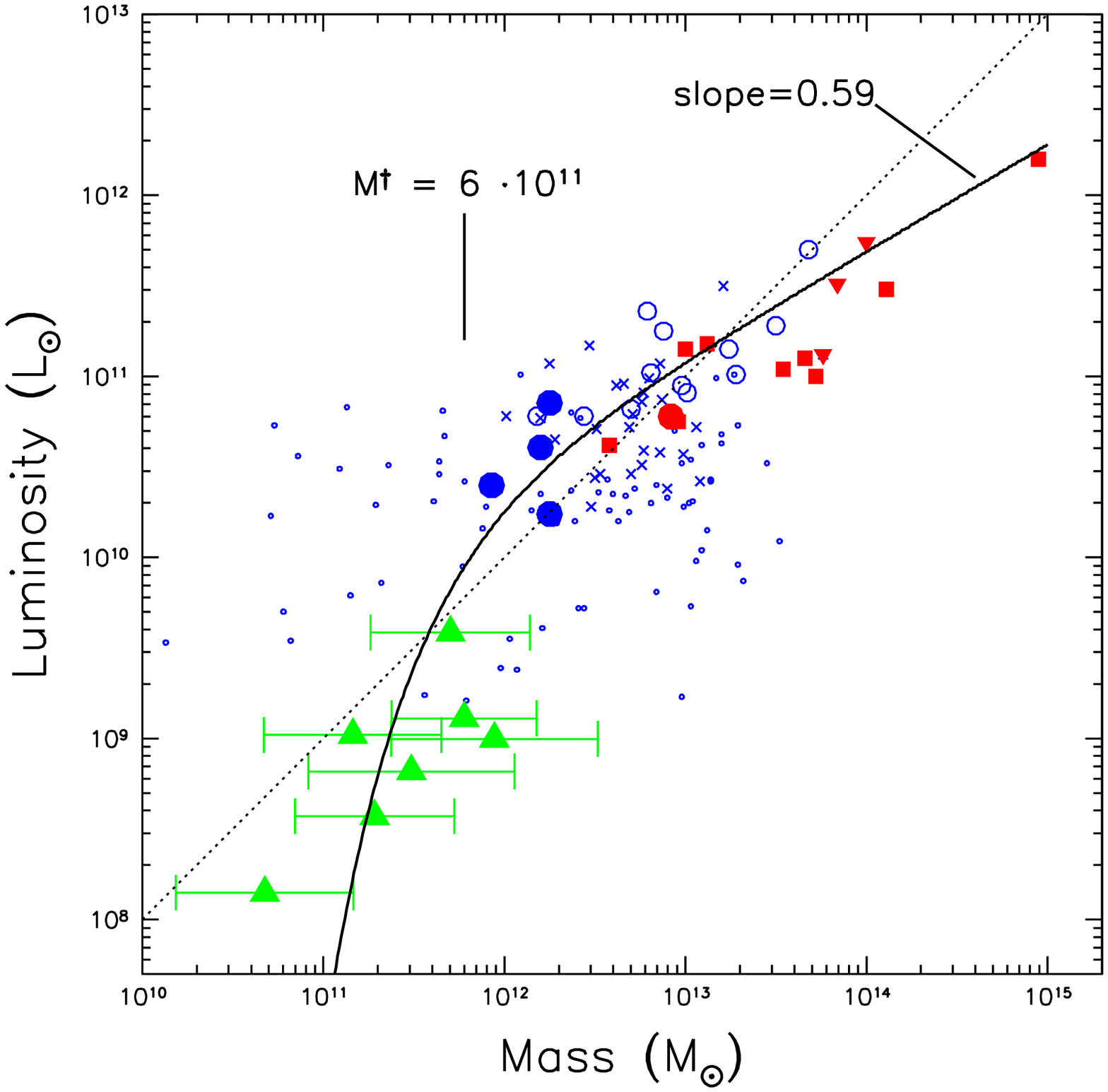}
\caption{Virial mass vs. blue light for groups of galaxies in the Local Supercluster.  Red: groups dominated by early types; blue: groups in the majority late types; green: associations consisting of only dwarf galaxies.  The dashed line illustrates $M/L_B = 100$.  The curve is described by Eq.~\ref{eq:ml}.}
\label{ml}
\end{center}
\end{figure}

Familiar groups can be roughly catagorized by whether the majority of luminous ($M_R < -17$) galaxies are elliptical/lenticular or spiral/irregular.  Those dominated by early morphological types are considered to be dynamically evolved while those dominated by late types, are considered to be at an early stage of development through merging.  The virial mass to blue light ratio, $M_{\rm v}/L_B$, of groups is strongly correlated with both the virial mass and the morphological content of groups as described by   \citet{2005ApJ...618..214T}.  Figure~\ref{ml} is an updated version of a figure in that reference.  The curve is a power law with a low mass cutoff:
\begin{equation}
L_B = 3.25 \times 10^{10}~M_{12}^{0.59}~{\rm exp}^{-0.6/M_{12}} .
\label{eq:ml}
\end{equation}
Rather than being described by a continuous curve, though, the distribution might be considered to consist of three regimes.  Groups dominated by early types typically have $M_{\rm v}/L_B$ values of several hundred in solar units.  Groups dominated by late type systems, like the MW and M31 groups, typically have $M_{\rm v}/L_B$ values below or near 100.  The prominence of young hot stars results in a lot of light per unit mass.  In the regime of low mass groups of dwarfs $M{\rm v}/L_B$ values can be very large, suggesting inefficiency in converting baryons into stars.

\begin{figure}[htbp]
\begin{center}
\includegraphics[scale=0.36]{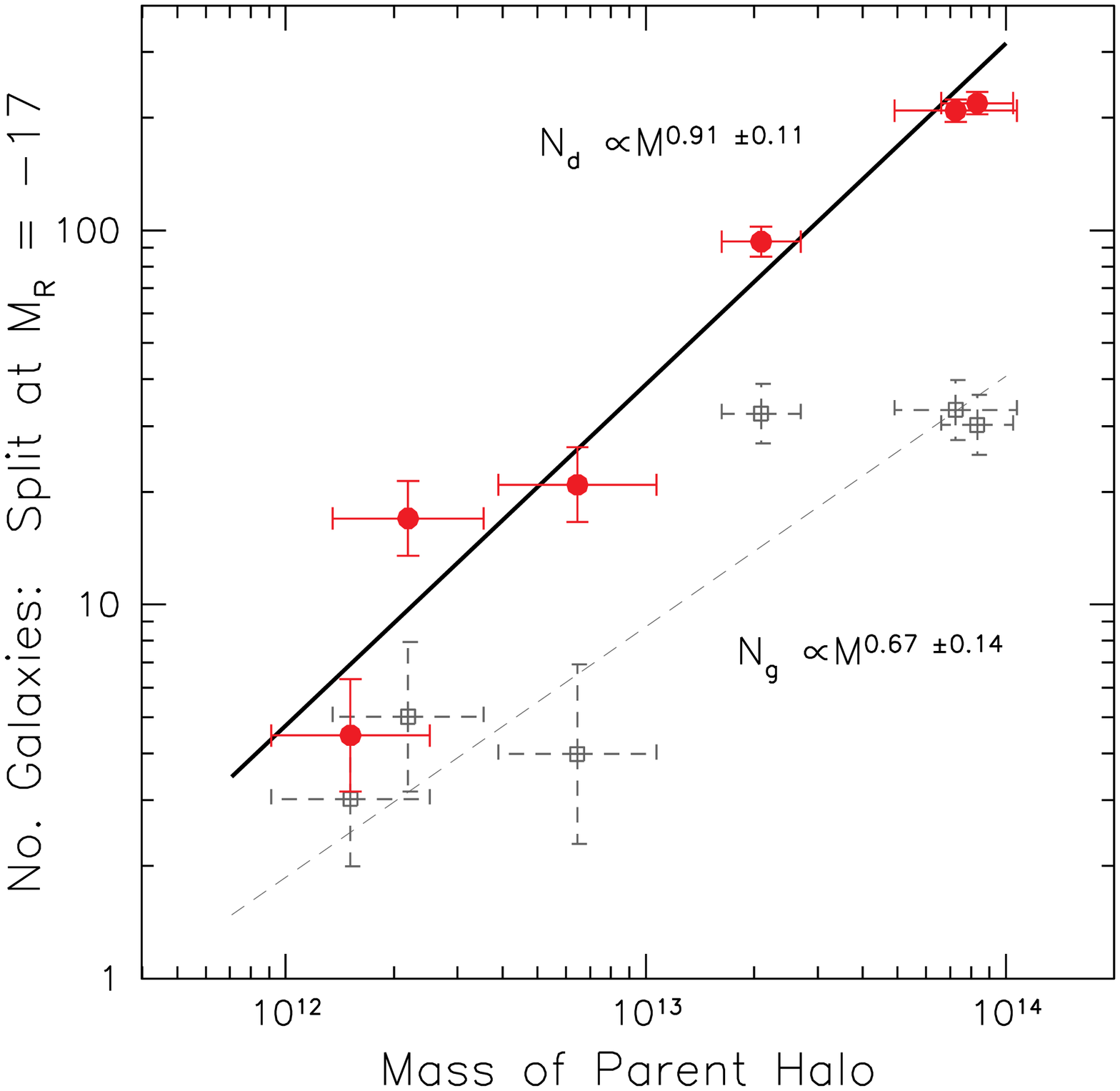}
\caption{Number of group members as a function of the mass of parent halo.  Red: only dwarfs with $-17 < M_R < -11$.  Grey: only giants with $M_R < -17$.  Solid line: correlation found with the dwarf sample: consistent with a constant number of dwarfs per unit parent halo mass.}
\label{Nd-M}
\end{center}
\end{figure}

Groups of both the early and late types were studied during the CFHT Megacam imaging campaign.  The completeness limit for membership studies was typically $M_R = -11$ or fainter.  In an effort to study the properties and possible variations in the luminosity function, galaxies were separated into two bins: `giants' with $M_R < -17$ and `dwarfs' with $-17 < M_R < -11$.  Small but significant variations in luminosity functions were found.  The variations could be characterized by variations in the ratio of dwarfs to giants.  The groups of predominantly early types have larger dwarf/giant ratios.  An interesting result is found if the number of giants or dwarfs is plotted against the group virial mass.  The correlation is poor with giants but very pronounced with dwarfs.  The result is shown in Figure~\ref{Nd-M}.  The straight line describes the relation
\begin{equation}
N_d = 5.2~M_{12}^{0.91 \pm 0.11}
\end{equation}
where $N_d$ is the number of dwarfs with $-17 < -M_R <-11$ in the group.  The slope is consistent within the errors with unity.  The number of dwarfs per unit halo mass is roughly constant.  A count of the number of dwarfs is a good measure of the mass of a group halo.

\section{Summary}

Halos can be observationally delineated by a density drop and a transition between dispersed orbits and infall at the radii of second turnaround, $r_{2t}$.  The scaling relations that are theoretically anticipated to exist over a wide range of halo masses are  confirmed from observations of nearby collapsed regions.  Specifically:

$\bullet$ Over the mass range from the halo of M31 to the Coma Cluster there is a correlation between projected second turnaround radius $R_{2t}$ and line-of-sight velocity dispersion $\sigma_p$:  $\sigma_p / R_{2t} = 368~h_{75}$~\kmsMpc.

$\bullet$ Over the same mass range, there is the correlation between $R_{2t}$ and the virial mass $M_{12}$ measured in units of $10^{12}~M_{\odot}$:  $R_{2t} = 0.215~M_{12}^{1/3}~h_{75}^{-2/3}$ Mpc.

$\bullet$ These two correlations can be combined to give the alternative, but not independent, relation:   $M_{\rm v} / M_{\odot} = 2.0 \times 10^6~\sigma_p^3~h_{75}^{-1}$.

In a few nearby groups distance information is sufficient to constrain the radius of first turnaround $r_{1t}$, the zero velocity surface between infall and cosmic expansion.  Three reasonably studied collapse regions define the relation $r_{1t} = 0.77~M_{12}^{1/3}~h_{75}^{-2/3}$~Mpc.  The ratio of first and second turnaround radii depends on the pace of the cosmic clock set by the matter-energy density of the universe.  At present the constraints are not tight but a tentative measurement favors $\Omega_{matter} \sim 0.15$ if the universe is topologically flat.

Among the groups large and small that have been discussed in some detail, a couple of particularly interesting details are worth highlighting.  The NGC~1407 Group has drawn attention because of the remarkably large blueshift of NGC 1400.  It is reported here that two additional members of the group, small dwarfs, now have measured relative blueshifts that are almost as extreme.  Then, given the attention that is being given to the discovery that M31 satellites lie in thin planes, it is noteworthy to see a related behavior in the galaxies around Centaurus~A.

Dwarf galaxies serve as markers of collapsed halos because they are numerous.  Dwarfs and giants alike that are located within the radii of second turnaround tend to be gas-poor `early' types while galaxies of all sizes outside the second turnaround tend to be gas-rich `late' types.  The transition between predominantly early and late populations is another indicator of the radius of second turnaround.

Dwarf galaxies serve as markers for collapsed halos in a way that is surprisingly quantitative.  If the number of dwarf galaxies with $-17 < M_R < -11$ are counted, then the number of these dwarfs in a halo depends linearly on the mass of the halo.  Count the number of dwarfs and one has a measure of the mass of the halo.

There are differences in the luminosity function of galaxies with environment \citep{2009MNRAS.398..722T}.  More dynamically evolved environments have somewhat steeper faint end slopes.  However, it appears that the reason is not more dwarfs per unit mass in evolved environments but fewer intermediate luminosity systems in the vicinity of $L^{\star}$, the luminosity that characterizes the exponential cutoff from the faint end power law distribution.  It is suspected that intermediate luminosity systems are being lost through mergers with the central dominant galaxy.  At faint luminosities, the production and depletion mechanisms are such that the number of dwarfs per unit halo mass remains roughly constant.

In a second paper in this series, the scaling relationships that have been defined in this study will be applied to the compilation of a group catalog of galaxies from the 2MASS redshift survey of \citet{2012ApJS..199...26H}.

\noindent
{\it Acknowledgements.} The research that has been described has been realized with the contributions of close collaborators Kristin Chiboucas, Brad Jacobs, Igor Karachentsev, Andi Mahdavi, Ed Shaya, and Neil Trentham.   
Special thanks to Gary Mamon for enlightening conversations about infall timing and caustics (Gary, please publish that paper on infall timing).
Ground-based observations have been made with the Canada-France-Hawaii, Keck, and Subaru telescopes.  Support over the many years leading to this paper has been provided by the US National Science Foundation and several awards from the Space Telescope Science Institute in connection with observations with Hubble Space Telescope.

\bibliography{paper}
\bibliographystyle{apj}

%\clearpage
%\begin{table}
%\vspace{-2cm}
%\includegraphics[scale=0.8]{pscollage1.ps}
%\caption{test \label{mycap}}
%\end{table}

\
\end{document}